\documentclass[aps,prb,reprint,groupedaddress,showpacs,showkeys]{revtex4-1}

\usepackage{graphicx}
\usepackage{subfigure}
\usepackage{amsmath}
\usepackage{color}
\usepackage{array} 
\usepackage[bold]{hhtensor}
\usepackage{geometry}

\begin{document}

\title[phase transition Ge]{Mesoscale computational study of the nano-crystallization of \\
amorphous Ge via a self-consistent atomistic - phase field coupling}

\author{C.~Reina$^{1,2}$, L.~Sandoval$^{1,3}$ and J.~Marian$^1$ }
\email[]{creina@seas.upenn.edu}
\address{$^1$Science and Technology Principal Directorate, Lawrence Livermore National Laboratory, Livermore, CA 94551, USA}
\address{$^2$Department of Mechanical Engineering and Applied Mechanics, University of Pennsylvania, Philadelphia, PA 19104-6315, USA}
\address{$^3$Theoretical Division T-1, Los Alamos National Laboratory, Los Alamos, NM 87545, USA}

\date{\today}

\begin{abstract}
Germanium is the base element in many phase-change materials, {\it i.e.}~systems that can undergo reversible transformations between their crystalline and amorphous phases. They are widely used in current digital electronics and hold great promise for the next generation of non-volatile memory devices. However, the ultra fast phase transformations required for these applications can be exceedingly complex even for single component systems, and a full physical understanding of these phenomena is still lacking.  In this paper we study nucleation and growth of crystalline Ge from amorphous thin films at high temperature using phase field models informed by atomistic calculations of fundamental material properties. The atomistic calculations capture the full anisotropy of the Ge crystal lattice, which results in orientation dependences for interfacial energies and mobilities. These orientation relations are then exactly recovered by the phase field model at finite thickness via a novel parametrization strategy based on invariance solutions of the Allen-Cahn equations. By means of this multiscale approach, we study the interplay between nucleation and growth and find that the relation between the mean radius of the crystallized Ge grains and the nucleation rate follows simple Avrami-type scaling laws. We argue that these can be used to cover a wide region of the nucleation rate space, hence facilitating comparison with experiments.
\end{abstract}


\maketitle

\graphicspath {
    {./}
    {./Figures/}
}

\section{Introduction}
The last two decades have seen an increasing interest in phase-changing materials (PCM), such as Ge-Sb-Te (GST), as promising candidates for next-generation nonvolatile memory devices \citep{Wuttig2005}. These materials display ultrafast and reversible phase transitions between their crystalline and amorphous states at high temperature, while displaying remarkable stability at low temperatures. Their potential originates from the striking differences in electrical and optical properties exhibited by each phase, which can reach several orders of magnitude in some instances \citep{Wuttig2007}. These two properties of GST materials ---reversible kinetics and high contrast in phase properties--- makes them ideal candidates for memory applications. Generally, a phase transformation is induced by some controlled external heat source that drives the system above the transition threshold. 

However, a complete understanding of these fast phase transformations at ultra high heating rates and their remarkable properties is still lacking. Even for single component systems such as Ge, the crystalline structures induced by laser-heating of amorphous thin films are very complex and depend on many factors such as the deposition history and morphology of the amorphous structure, the intensity and duration of the heating pulse, and the substrate structure, among others \citep{Nikolova2013,Peng1997,Sun2007}. The interplay between all these processes governs the kinetic and thermodynamic evolution of PCM, which must be understood in terms of the microscopic properties of each phase. Modeling and simulation can provide physical insight into the governing phenomena, as well as a path to understanding the physical evolution of PCM under the conditions just discussed.

The goal of this paper is two-fold: (i) to present a novel multiscale model for phase transformations in a heat bath at constant temperature; and (ii) to obtain scaling laws for Ge-crystal mean radius evolution as a function of the nucleation rate during the recrystallization process. Our approach consists of a thermodynamically consistent phase field model (TCPF) that reproduces \emph{exactly} the interface energetics and kinetics of atomistically computed crystallization fronts. As an additional feature, the interface thickness may be chosen arbitrarily large while preserving this exact atomistic-to-continuum coupling, thus delivering a highly efficient multiscale computational model.

The paper is organized as follows. We begin in Section \ref{Sec:PhysicalModel} by discussing the physical processes captured by the model (nucleation, growth, role of anisotropy, etc.) and the boundary conditions employed. In Section \ref{Sec:Methods} we describe the mathematical formulation of the TCPF model, and prove that the energetics and kinetics of planar fronts at constant temperature obey universal relations when expressed in appropriate non-dimensional units. These relations lead to a direct correspondence between the phase field parameters and physical quantities that can be measured via atomistic simulations, and thus provide with an accurate coupling strategy between the two scales. Additionally, in the same section, we present detailed information on the evaluation of continuum data from atomistic calculations, namely, the free energies for each of the faces, the interface velocities and the transformation pathway between the amorphous and crystalline phase. Next, in Section \ref{Sec:Results} we show the results of the atomistic calculations as a function of temperature and interface orientation, compute the invariance solutions of the Allen-Cahn equations and obtain the related phase field parameters. Following these results, we use the developed multiscale model to analyze, in Section \ref{Sec:Discussions}, the interplay between nucleation and growth during crystallization kinetics under isotropic and anisotropic growth conditions. We conclude the paper in Section \ref{Sec:Summary}, with a summary of the results.

\section{Physical model} \label{Sec:PhysicalModel}
When an amorphous Ge thin film is heated above the glass transition temperature ($T_g$) but below the melting temperature ($T_m$), it spontaneously crystallizes. The difference in free energies between the two phases constitutes the driving force for crystallization, which decreases with temperature. At the same time, the mobilities and nucleation rates rapidly increase as the temperature rises. For most PCMs, there is an intermediate temperature at which an optimal compromise is achieved, and the crystallization rate is maximized. For Ge, such temperature is around $1100$ K, cf.~Sec. \ref{Sec:Results_MD}, which can be achieved, for instance, by high-intensity pulsed laser heating \citep{Nikolova2010}. Under those conditions, crystals nucleate and grow until full impingement is achieved. Complete crystallization occurs over tens of nanoseconds, and the resulting crystals have a mean radius of tens of nanometers \citep{Nikolova2010}.

Experiments suggest that nucleation occurs homogeneously in space throughout the thin film with no preferential texture, which typically indicates heterogeneous nucleation at uniformly distributed imperfections and/or irregularities in the amorphous microstructure. From a theoretical point of view, nucleation is ill-defined despite substantial efforts \citep{Stiffler1990,Lu1996,Schwen2013}, and here the nucleation rate is simply taken taken as a parameter of the model, which can be changed to understand its effect on the overall crystallization process.

Once crystals have nucleated, they grow by front propagation in the surrounding amorphous material, or, if in contact with another crystal, by grain boundary motion. In this paper, we focus solely on the former, and leave the study of coarsening of the recrystallized nanograined material for future studies. 

The mesoscopic time and length scales involved in the crystallization process (microns and tens of nanoseconds) are prohibitive for fully atomistic simulations. Hence, in this paper we adopt a computational approach based on the phase field equations parameterized with atomistic data. The details of the method are described in the following section. In the remainder of the paper it will be assumed that the film is thin enough for critical nuclei to span the full thickness and for the problem to be treated as two-dimensional. Furthermore we will consider that the amorphous thin film is free-standing and capable of fully releasing stresses via bending.

\section{Methods} \label{Sec:Methods}

\subsection{The phase field model}\label{phase}

The phase-field method is a computational approach widely used in modeling morphological and microstructural evolution in materials \citep{Chen2002}, including solidification \citep{Boettinger2002}, fracture \citep{Miehe2010}, dislocation dynamics \citep{Koslowski2002}, crystal nucleation \citep{Granasy2007} and growth \citep{KrillChen2002}, and diffusion processes \citep{Cahn1997}, among others. The method uses a so-called phase-field variable, $\phi$, to distinguish among the different material phases and implicitly track interface motion. For example, in this work we take $\phi = 1$ to represent the crystalline phase, and $\phi = 0$ for the amorphous region. The interface between the two phases is then described as a smooth transition of the variable $\phi$ through a small thickness. 

In phase field models, the equilibrium configuration of a material point in homogeneous state is assumed to be characterized by a Gibbs free energy $g(\phi, T, p,\ldots)$ that depends on the phase field variable and a set of thermodynamic state variables such as temperature and pressure. The total free energy of the system is then described via a Landau-Ginzburg free energy functional of the form \citep{Penrose1990}
\begin{equation} \label{Eq:TotalEnergy}
\mathcal{G} = \int_{\Omega} \Big[ g(\phi, T, p, \ldots) + \frac{m^2(\theta, T,p,\ldots)}{2} (\nabla \phi)^2 \Big] d\Omega,
\end{equation}
where the second term accounts for the additional contribution due to the presence of the interface, which is characterized by a non-vanishing value of $\nabla\phi$. The variable $m$ in Eq.~\ref{Eq:TotalEnergy} sets the interfacial energy and it is a function of the state variables and the misorientation $\theta$, defined as the relative angle between the crystal lattice $\theta_C$ and the normal to the interface, i.e. $\theta = \theta_C - \arccos(\frac{\nabla \phi}{|\nabla \phi|})$. The existence of a total free energy of the form just described is a standard postulate in transport processes in continuum mechanics \citep{CahnHilliard1958}. For the system of interest in this paper, we will assume in the following that the only thermodynamic variable dependence is with the temperature, which is assumed to be controlled by thermostatting to a heat bath.

The system's time evolution is then modeled using Allen-Cahn's equation \citep{AllenCahn1979, Chen2002}
\begin{equation} \label{Eq:General_PF}
\begin{split}
\frac{\partial \phi}{\partial t} &= - M(T,\theta)\frac{\delta \mathcal{G}}{\delta \phi} \\
&= M(T,\theta) \Bigg[ m^2(T,\theta)\nabla^2 \phi-\left(\frac{\partial g}{\partial \phi} \right)_{T} \Bigg],
\end{split}
\end{equation}
where $M > 0$ is the mobility\footnote{In the derivation of this equation it has been neglected the dependence of $\theta$ on $\nabla \phi$}.  This simple form of the kinetic relation at constant temperature guarantees a decrease of total free energy of the system $\mathcal{G}[T,\phi]$ along the solution path, in accordance to the second law of thermodynamics.

We further assume that the local free energy for intermediate values of $\phi$ can be computed from the amorphous free energy $g_A(T)$ and the crystalline one $g_C(T)$ as
\begin{equation} \label{Eq:local_free_energy}
g(T,\phi) = g_A(T) - q(\phi) \Delta g(T) + B(T) g_{dw}(T,\phi),
\end{equation}
with
\begin{equation}
\begin{split}
q(\phi) & = \phi^3(10-15\phi+6\phi^2), \\
g_{dw}(T,\phi) &= 16 \phi^2 (1-\phi)^2, \\
\Delta g(T)& = g_A(T) - g_C(T).
\end{split}
\end{equation}

$\Delta g(T) > 0$ is the driving force for the transformation and $B(T)$ represents the energy barrier along the transformation path $\left(B(T) = g\left(T,\phi=1/2 \right) -\frac{g_A(T)+g_C(T)}{2} \right)$. This functional form guarantees that the free energy has two local minima at all temperatures at $\phi=0$ and $\phi=1$, which correspond respectively to the pure amorphous and crystalline phases. The kinetic equation (\ref{Eq:General_PF}) can then be written as
\begin{equation}
\begin{split}
\frac{1}{M(T,\theta)} \frac{\partial \phi}{\partial t} &= m^2(T,\theta) \nabla^2 \phi - B(T) g'_{dw}(\phi) \\
&\phantom{=}+ \Delta g(T) q'(\phi),
\end{split}
\end{equation}
or equivalently, as
\begin{equation} \label{Eq:phi_constantT}
\begin{split}
\alpha(T,\theta) \epsilon^2(T,\theta) \frac{\partial \phi}{\partial t} &= \epsilon^2(T,\theta) \nabla^2 \phi - g'_{dw}(\phi) \\
&\phantom{=}+ \Delta \bar{g}(T) q'(\phi),
\end{split}
\end{equation}
where
\begin{align} \label{Eq:epsilon}
&\epsilon^2(T,\theta) := \frac{m^2(T,\theta)}{B(T)},\\ \label{Eq:alpha}
&\alpha(T,\theta) := \frac{1}{M(T,\theta) m^2(T,\theta)},\\
&\Delta \bar{g} (T) := \frac{\Delta g(T)}{B(T)},
\end{align}
and $\epsilon$ has units of length, $\alpha$ has units of an inverse diffusivity, and $\Delta\bar{g}$ is non-dimensional.
Equation (\ref{Eq:phi_constantT}) determines the growth of existing nuclei. Nucleation, for its part, is treated explicitly in this work by introducing cylindrical critical nuclei in the simulation with random orientation and with a probability that is consistent with a certain nucleation rate. It is assumed that the nuclei span the entire thickness of the film, so that their critical radius $R_c(T)$ in 2D can be estimated via classical nucleation theory as
\begin{equation} \label{Eq:CriticalRadius}
R_c(T)  = \frac{2 g_{int}(T,\theta)}{\Delta g(T)},
\end{equation}
where $g_{int}$ is the interfacial free energy.

\subsection{Invariance relations of the phase field equation}

\subsubsection{Steady state velocity}
The evolution of a flat crystalline-amorphous interface at constant temperature obeys an invariance relation that we proceed to characterize.  To that end, consider a set of parameters $\epsilon$ and $\alpha$ and the corresponding ones of a reference problem ($\epsilon_0$, $\alpha_0$). The solutions characterizing the motion of the front for both sets of parameters are denoted by $\phi$ and $\phi_0$ respectively, and are given by Eq.~(\ref{Eq:phi_constantT})
\begin{equation} \label{Eq:phi_phi0_1}
\begin{split}
\alpha \epsilon^2 \frac{\partial \phi}{\partial t} &= \epsilon^2 \nabla^2 \phi - g'_{dw}(\phi) + \nabla\bar{g}\ q'(\phi), \\
\alpha_0 \epsilon^2_0 \frac{\partial \phi_0}{\partial t} &= \epsilon_0^2 \nabla^2\phi_0 - g'_{dw}(\phi_0) + \nabla\bar{g}\ q'(\phi_0).
\end{split}
\end{equation}

These equations lead to a planar front moving at constant velocity in steady state, that we call $v$ and $v_{0}$ respectively. With the change of variables
\begin{equation}
\begin{split}
&\phi(x,t) = \phi(x-vt) = \phi^*(x^*),\\
&\phi_0(x,t) = \phi_0(x-v_{0}t) = \phi_0^*(x^*),
\end{split}
\end{equation}
Eqs.~(\ref{Eq:phi_phi0_1}) transform into the following ordinary differential equations
\begin{equation}
\begin{split}
-\alpha \epsilon^2 v \nabla^* \phi^* &= \epsilon^2 {\nabla^*}^2 \phi^* - g'_{dw}(\phi^*) + \nabla \bar{g}\ q'(\phi^*), \\
-\alpha_0 \epsilon_0^2 v_0 \nabla^* \phi_0^* &= \epsilon^2_0 {\nabla^*}^2 \phi_0^* - g'_{dw}(\phi^*_0) + \nabla \bar{g}\ q'(\phi_0^*),
\end{split}
\end{equation}
where $\nabla^*$ represents the gradient with respect to coordinates $x^*$. If we now define
\begin{equation} \label{Eq:VariableChange}
y^* = \frac{\epsilon_0}{\epsilon}x^*
\end{equation}
and denote $\phi^*(x^*)=\phi^* \left( \frac{\epsilon}{\epsilon_0}y^*\right) = \phi^{**}(y^*)$, then the two equations become
\begin{equation}
\begin{split}
-\alpha \epsilon \epsilon_0 v \nabla^{**} \phi^{**} &= \epsilon^2_0 {\nabla^{**}}^2 \phi^{**} - g'_{dw}(\phi^{**}) \\
&\phantom{=} + \nabla \bar{g}\ q'(\phi^{**}) \\
-\alpha_0 \epsilon_0^2 v_{0} \nabla^* \phi_0^* &= \epsilon^2_0 {\nabla^*}^2 \phi_0^* - g'_{dw}(\phi^*_0) \\
&\phantom{=}+ \nabla \bar{g}\ q'(\phi_0^*),
\end{split}
\end{equation}
which are identical for $\alpha_0\epsilon_0v_{0} = \alpha \epsilon v$. In other words, the non-dimensional steady state velocity ($\alpha \epsilon v$) satisfies an invariance relation of the form
\begin{equation} \label{Eq:v_int}
\alpha(T,\theta) \epsilon(T,\theta)v(T,\theta)= I_v\left( \Delta \bar{g}(T,\theta)\right)
\end{equation}

\subsubsection{Interfacial energy}
Similarly, the non-dimensional interface energy $\left(\frac{g_{int}(T,\theta)}{B(T)\epsilon(T,\theta)} \right)$ can be shown to satisfy
\begin{equation} \label{Eq:f_int}
\frac{g_{int}(T,\theta)}{B(T,\theta)\epsilon(T,\theta)} = I_g(\Delta \bar{g}(T,\theta)).  
\end{equation}
This relation can be obtained by defining the interface energy as the difference between the sum of energies of the bulk systems and the energy of the two-phase configuration separated by a sharp interface. Choosing a reference frame that moves with the interface steady state velocity, the surface energy reads
\begin{equation} \label{Eq:gint}
\begin{split}
g_{int}(T,\theta) &= \int_{\ell} \Big[\frac{m^2(T,\theta)}{2}(\nabla^* \phi^*)^2 + B(T) g_{dw}(\phi^*) \\
&\phantom{=}- \Delta g(T) q(\phi^*) + g_A(T)\Big] dx^* \\
&\phantom{=}- \frac{h_{\ell}}{2}(g_A(T)+g_C(T)),
\end{split}
\end{equation}
where $\ell$ is the range of the coordinate orthogonal to the interface and $h_{\ell} =|\ell|$ represents the system length in such direction. Without loss of generality, we consider that the interface is at the center of the domain under consideration. Then, at constant temperature,
\begin{equation} 
\begin{split}
g_{int}(T,\theta) &=m^2(T,\theta) \Bigg[ \int_{\ell} \frac{(\nabla^* \phi^*)^2}{2} dx^*\Bigg]  \\
&\phantom{=}+ B(T) \Bigg[ \int_{\ell}g_{dw}(\phi^*)dx^*\Bigg] \\
&\phantom{=}+ \Delta g(T) \Bigg[\frac{h_{\ell}}{2} - \int_{\ell}q(\phi^*)dx^*\Bigg].
\end{split}
\end{equation}

By recourse to the changes of variable expressed by Eq.~(\ref{Eq:VariableChange}) and the equivalence $\phi^{**}(y^*)=\phi^*_0(x^*)$, this equation can be rewritten as
\begin{equation} 
\begin{split}
\frac{g_{int}(T,\theta)}{B(T) \epsilon(T,\theta)} &=\epsilon_0 \Bigg[ \int_{\ell_0} \frac{(\nabla^* \phi^*_0)^2}{2} d x_0^*\Bigg]  \\
&\phantom{=}+ \frac{1}{\epsilon_0} \Bigg[ \int_{\ell_0}g_{dw}(\phi^*_0) d x_0^*\Bigg] \\
&\phantom{=}+ \frac{\Delta \bar{g}(T,\theta)}{\epsilon_0} \Bigg[\frac{h_{\ell_ 0}}{2} - \int_{\ell_0}q(\phi^*_0) d x_0^*\Bigg] \\
&= I_g(\Delta \bar{g}(T,\theta))
\end{split}
\end{equation}
and the sought-after result is obtained.

\subsection{Coupling strategy}

The phase field model at constant temperature described by Eq.~(\ref{Eq:phi_constantT}) requires that the following magnitudes be defined: $g(T,\phi)$, $M(T,\theta)$, $m(T,\theta)$ and $B(T)$.
In light of the invariance relations previously derived, these parameters can be obtained directly via atomistic simulations of the following kind:
\begin{enumerate}
\item[(a)] Free energy calculations using thermodynamic integration of bulk crystalline and amorphous phases at zero pressure and constant temperatures. These yield $g_C(T)$ and $g_A(T)$.
\item[(b)] Transition energy barrier between the amorphous and crystalline states at zero pressure and constant temperature. To obtain $B(T)$ we equate the transition path to an activated process described by a general configurational nonlinear many-body reaction coordinate\cite{carter1989}.
\item[(c)] Free energy calculations at zero pressure and constant temperature to compute $g_{int}(T,\theta)$. As well, molecular dynamics simulations of moving fronts in steady state to extract interface velocities $v(T,\theta)$.
\end{enumerate}

Equations (\ref{Eq:v_int}) and (\ref{Eq:f_int}) allow to explicitly relate the atomistic data with the phase field parameters $M(T,\theta)$ and $m(T,\theta)$ as
\begin{align} \label{Eq:m}
&m(T,\theta) = \frac{g_{int}(T,\theta)}{\sqrt{B(T)} I_g(\Delta \bar{g})}, \\ \label{Eq:tau}
&\tau(T,\theta) = \frac{1}{M(T,\theta)} = \frac{g_{int}(T,\theta)}{v(T,\theta)} \frac{I_v(\Delta \bar{g})}{I_g(\Delta \bar{g})}.
\end{align}
The details of such computations for the different subsystems are described in the following section.

\subsection{Free energy calculations from atomistic simulations} \label{subsec:bulk}
To compute the free energies of bulk phases we use thermodynamic integration. The Helmholtz free energy\footnote{In our problem stresses are assumed to be zero, and therefore the Helmholtz free energy equals the Gibbs free energy at all temperatures.} is defined as
$$F=E-TS,$$
where $E$ and $S$ are the internal energy and the entropy.
The approach employed here to compute $F$ for solid systems\footnote{Defined as those where self-diffusion and/or self-relaxation processes occur on time scales beyond the thermal equilibration time of the system.} involves two steps. First, a free energy $F_0$ is obtained at a temperature $T_0$ in each case. Second, the internal energy is computed for the desired range of temperatures as
$$E=\langle V+K\rangle,$$
where $V$ and $K$ are the potential and kinetic energies, and, under the usual assumption of ergodicity, $\langle\cdot\rangle$ denotes time average. The free energy at a given temperature $T$ can then be obtained by solving the Gibbs-Helmholtz integral
\begin{equation}
\frac{F}{T}=\frac{F_0}{T_0}-\int_0^T\frac{E(\tau)}{\tau^2}d\tau,
\label{gh}
\end{equation}
where $E(T)$ is made to be a smooth integrable function that expresses the temperature dependence of the internal energy. 

To obtain $F_0$, we employ the $\lambda$-integration technique. The use of this technique is well documented in the literature \cite{lambada1,lambada2} and here we only provide a brief overview. It is assumed that the system at $T_0$ is represented by the following energy function
\begin{equation}
\tilde{U}(\lambda)=(1-\lambda)^kU'+\lambda^kU,
\label{lambada}
\end{equation}
where $k\ge0$ is a constant, and $U(\vec{r})$ and $U'(\vec{r})$ are, respectively, the potential energy functions of the system under study and a reference system whose free energy is known. For practical reasons, $U'$ is typically taken to be the potential energy function of an analytical or a numerical system of coupled harmonic oscillators. 
From here, $F_0$ is straightforwardly obtained by integrating along a $\lambda$ trajectory between 0 and 1
\begin{equation}
\begin{aligned}
F_0(U)&=F_0(U')~+&\\
&+\int^1_0k\left[\lambda^{k-1}\langle U\rangle-(1-\lambda)^{k-1}\langle U'\rangle\right] d\lambda.&
\end{aligned}
\label{fu}
\end{equation}
In this work, the calculation hypotheses warranted the use of linear thermodynamic integration ($k=1$). The numerical behavior of Eq.\ (\ref{fu}) under other choices of $k$ has been reviewed in the literarure \cite{resat1993}, and we refer the reader to those works for more details.

\begin{figure}[h]
        \centering
                \includegraphics[width=1.0\linewidth]{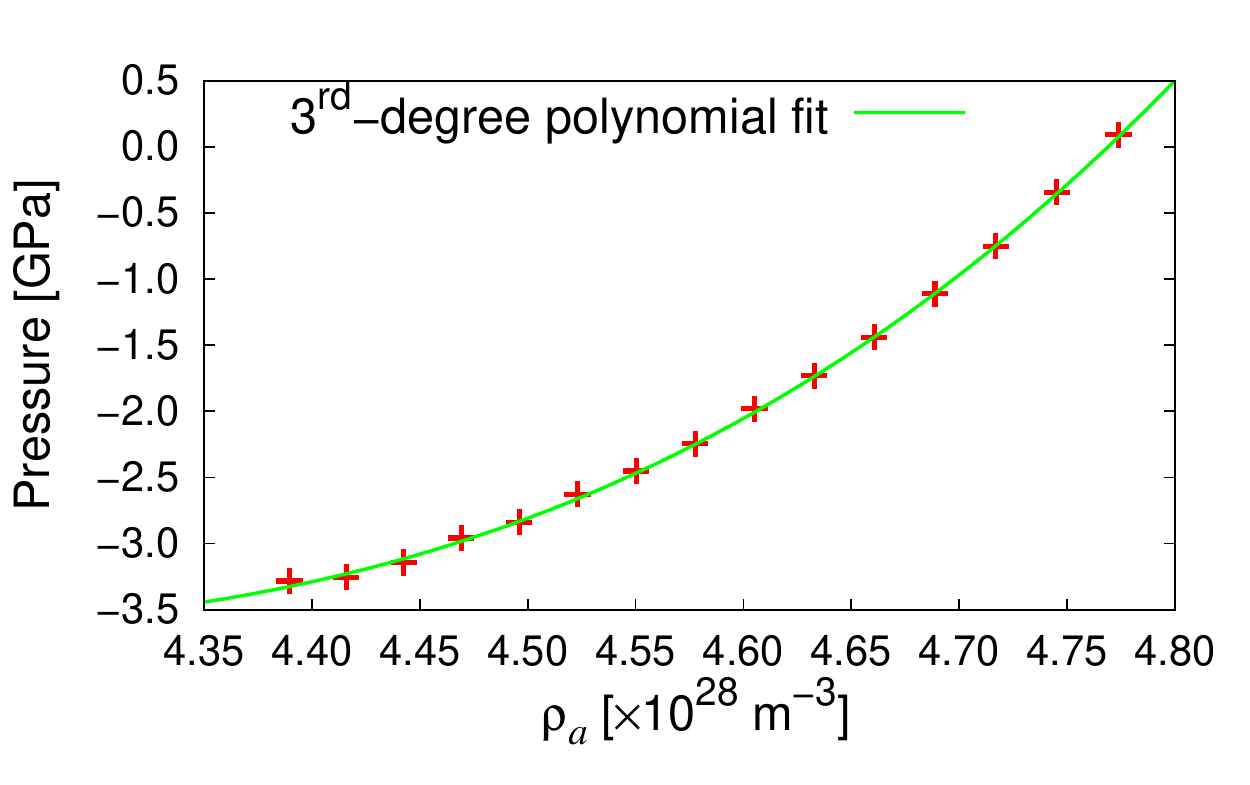}
         \caption{Equation of state for liquid Ge at 1500 K.}\label{fig:ideal}
\end{figure}

When internal transport processes derived from the existence of viscous flow, {\it e.g.}~as in liquids, are important, Eq.~(\ref{fu}) can no longer be used in conjunction with a reference harmonic crystal to compute $F_0$. In such cases, the free energy is obtained as \cite{glosli1999,frenkelbook}
\begin{equation}
\begin{aligned}
F_0&=F_{g}(\rho_0,T_0)+&\\
&+\int_0^{\rho_0}d\rho\left[\frac{p(T_0,\rho)-\rho k_BT_0}{\rho^2}\right],&
\end{aligned}
\label{liquid}
\end{equation}
where $p$ is the pressure, $\rho_0$ is the system's density at a temperature $T_0$ above the supercritical temperature (here we have taken $T_0=1500$ K) and $F_{g}(\rho,T)$ is the free energy of an ideal gas at density and temperature $\rho$ and $T$ \cite{brou1997}.
The integrand in the second term of the r.h.s.~of Eq.~(\ref{liquid}) represents an isothermal expansion from $\rho_0$ to zero density ({\it i.e.}~infinite volume), where the system effectively behaves as an ideal gas. This expansion should be reversible, which means that no first-order transition ---e.g.~liquid-gas--- should be traversed. For its part, the equation of state $p(T_0,\rho)$ is obtained from a set of canonical ensemble calculations of liquid Ge at $T_0$. This is shown in Fig.~\ref{fig:ideal}. The resulting data are fitted to a 3$^{\rm{rd}}$ degree polynomial and the integral in Eq.~(\ref{liquid}) is solved to yield the free energies.
\begin{figure}[h]
        \centering
                \includegraphics[width=1.0\linewidth]{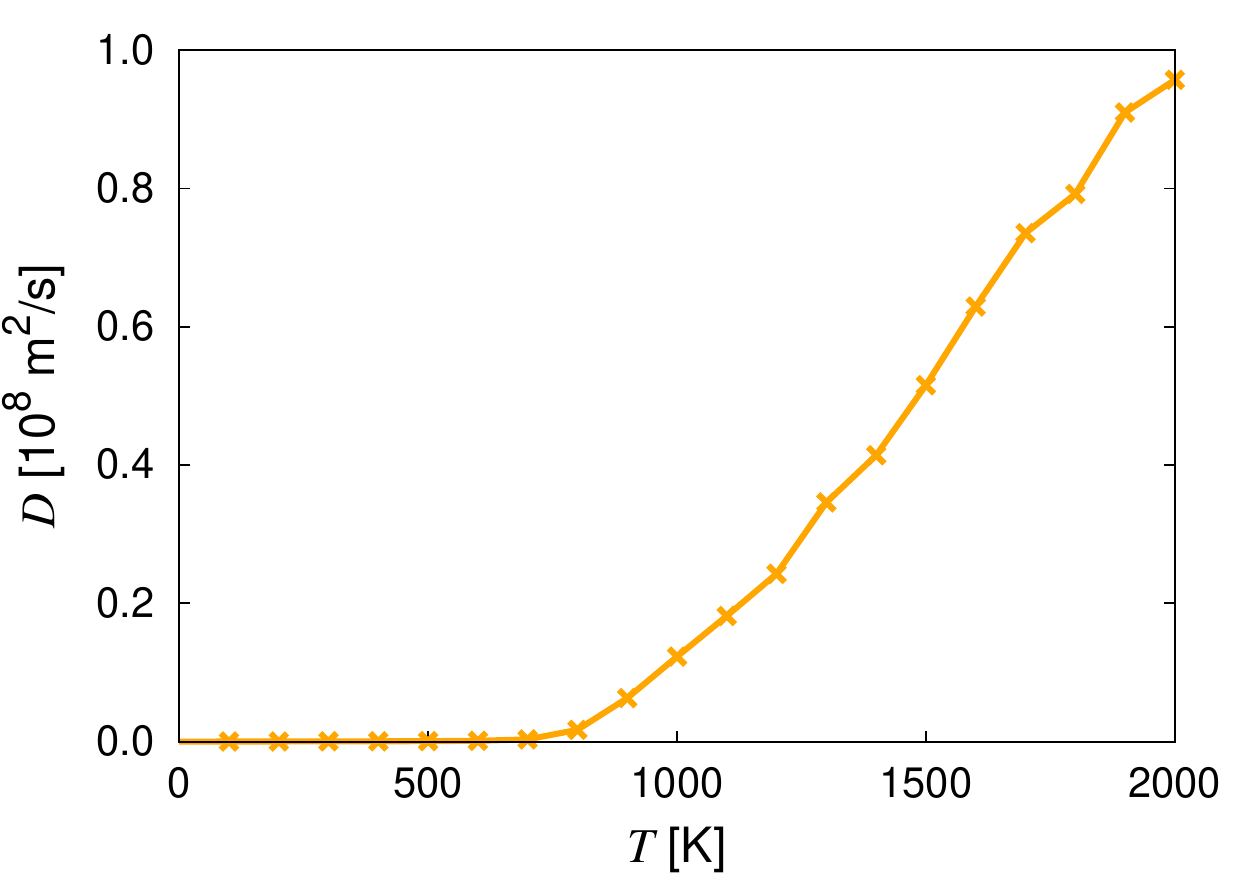}
         \caption{Self-diffusivity of amorphous Ge as a function of temperature.}\label{fig:diff}
\end{figure}	

To establish the limits of applicability of Eqs.~(\ref{fu}) and (\ref{liquid}), we compute the self-diffusivity of amorphous Ge as a function of temperature. Results are shown in Figure \ref{fig:diff}. Our calculations clearly demonstrate that internal transport is important only above 900 K, at which point Eq.~(\ref{liquid}) is the pertinent one for computing free energies.  

\subsection{Calculations of interfacial free energies}
For ideal (perfectly-flat, stress-free) interfaces, the interfacial free energy $\gamma$ can be defined as the change in free energy due to an increase in interface surface area $A$
$$\gamma=\left(\frac{\partial F}{\partial A}\right)_{N,p,T}.$$
From a practical standpoint, interfacial free energies are computed using the Dupr\'e equation
\begin{equation}
\gamma_{ij}=\gamma_i+\gamma_j-W_{ij},
\label{dupre}
\end{equation}
where $\gamma_i$ is the surface free energy of solid $i$ in a vacuum and $W_{ij}>0$ is the so-called work of \emph{adhesion}. The above formula results from a two-step construction in which (i) slabs with exposed area $A$ are created in a vacuum from bulk systems $i$ and $j$, and (ii) these new surfaces are brought into contact, releasing an amount of work $W_{ij}$. 
Using the bulk free energy densities, $f_i$, computed in Section \ref{subsec:bulk} by thermodynamic integration, we can express Eq.~(\ref{dupre}) as
$$\gamma_{ij}=\frac{F_{ij}(\Omega_{ij})-\left(f_i\Omega_i+f_j\Omega_j\right)}{2A},$$
where $F_{ij}$ is the free energy of a system with total volume $\Omega_{ij}=\Omega_i+\Omega_j$ of a simulation cell containing a volume $\Omega_i$ of phase $i$ and a volume $\Omega_j$ of phase $j$ separated by the interface. As for the bulk phases, $F_{ij}$ is computed using $\lambda$-integration. The calculation is simplified due to the fact that both phases correspond to monoatomic solid systems and thus the same reference thermodynamic system can be used for both.
Because cohesion in crystalline solids is anisotropic --while in amorphous solids or liquids is isotropic--, we have repeated this procedure for a number of different surface orientations to assess the orientation dependence of the interfacial free energy. 

As a final point, we note that, in reality, atomistic interfaces are intrinsically rough and accurate methods for computing interfacial free energies that account for such atomic-scale roughness have been developed \cite{sun2004}.

\subsection{Molecular dynamics calculations of interface mobility}

To calculate interface mobilities, large-scale MD simulation of crystal-amorphous mixing are performed. For this, a bicrystal with the desired interface orientation is constructed and the system is left to evolve at a given temperature. The temperature must be kept constant on the boundaries of this box using some suitable themostat. It is important to ensure that no temperature control is applied in the interface region, as this could result in thermodynamic artifacts. 
Following Hoyt and Song \citep{song2012} we measure the rate of change in potential energy as the front advances, $dU/dt$, and obtain the velocity as
$$v=-\frac{1}{2A\Omega h_l}\frac{dU}{dt},$$
where $A$ is again the interface surface area, $\Omega$ is the volume per atom in the crystalline phase at the temperature of interest, and $h_l$ is the latent heat, which in this case is simply equal to the internal energy difference between the two phases. As above, the factor of two in the above expression reflects the use of periodic boundary conditions, which introduces two interfaces in the simulation volume.

\subsection{Interatomic potential and boundary conditions}
Next, we discuss the most salient features of the potential energy function $U(\vec{r})$ employed here.
We use a Stillinger-Weber potential parameterized by Posselt and Gabriel for Ge\cite{posselt2009}. The potential builds on the parameterization by Ding and Anderson\cite{ding1986}, which yields the correct values for the cohesive energy and lattice constant. The elastic constants, melting point, as well as the formation and migration energies of point defects for diamond-structure Ge are reproduced within certain limits. The energetics of other crystalline phases and the structure of the liquid are described reasonably well.

All the simulations were run in the isobaric-isothermal ensamble $NpT$ --where $N$ is the number of particles, $p$ is the pressure,and $T$ the abolute temperature-- using periodic boundary conditions in three dimensions. The pressure and temperature were controlled using a Nos\'e-Hoover thermostat with a damping constant of 0.1 ps in both cases.

\section{Atomistic results and parametrization of the phase field model} \label{Sec:Results}
\subsection{Molecular dynamic results} \label{Sec:Results_MD}
\begin{figure}[ht]
\centering
\subfigure[]{
 {\includegraphics[width=0.48\textwidth]{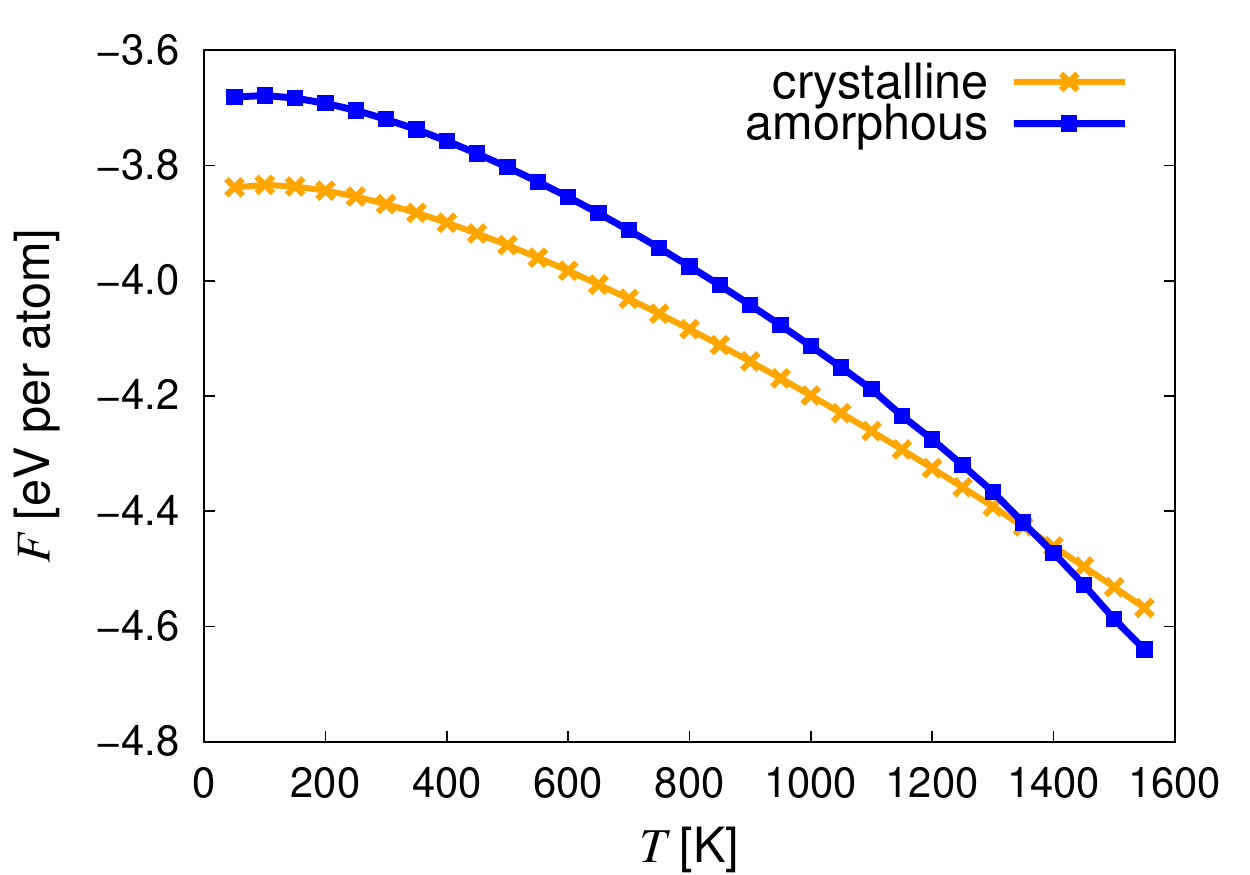}}
    \label{Fig:FreeEnergy_Temperature}
}
\subfigure[ ]{
  {\includegraphics[width=0.48\textwidth]{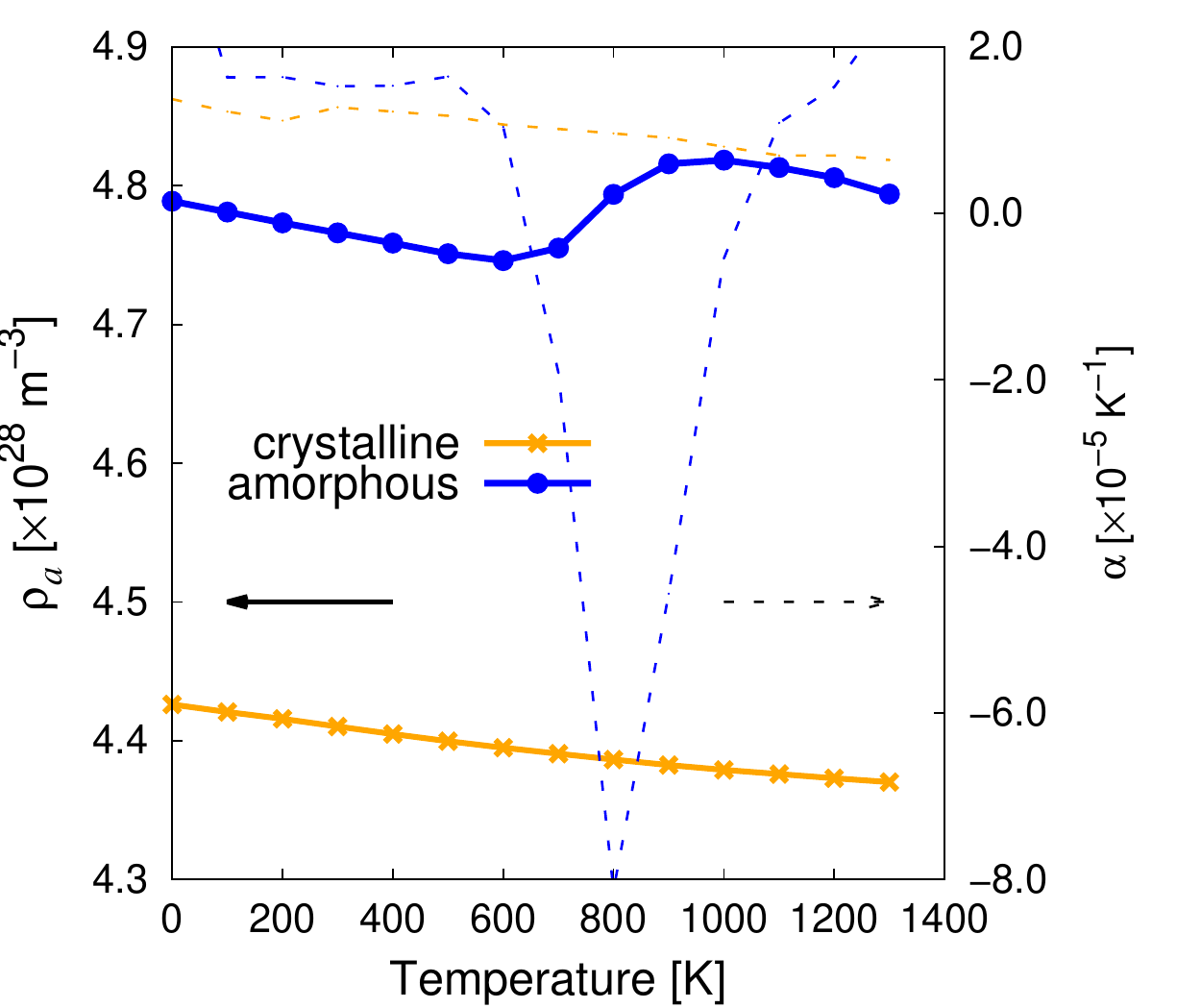}}
    \label{Fig:Volume_Temperature.pdf}
}
\caption{Results of single phase atomistic simulations at zero pressure: \subref{Fig:FreeEnergy_Temperature} free energy and \subref{Fig:Volume_Temperature.pdf} atomic density as a function of temperature.}
\label{Fig:FreeEnergy_LatticeConstant}
\end{figure}

First, we calculate the free energy of the crystalline and amorphous phases as a function of temperature. Figure \ref{Fig:FreeEnergy_Temperature} shows the results for each case, with the crystalline phase being the more stable one up to a temperature of 1350 K. This value can be considered as the melting point for the numerical force field $U(\vec{r})$ employed here, about 10\% larger than the experimental value of 1210 K \cite{hall1955}. Figure \ref{Fig:Volume_Temperature.pdf} shows the variation of the atomic density $\rho_a$ and the thermal expansion coefficient $\alpha_t$ with of temperature. As the figure shows, the amorphous structure is always approximately 10\% denser than the crystalline phase, and is seen to undergo a smooth transition at $\approx$800 K. This temperature correlates well with the point at which the self diffusivity is markedly nonzero (cf.~Fig.~\ref{fig:diff}), and can be assigned to the reversible liquid-glass transition. The temperature at which we have recorded this transition is in excellent agreement with experimental measurements of the glass transition temperature for Ge, $T_g\approx810$ K \cite{angell2000}. This glass-liquid transition manifests itself as a marked dip in the value of $\alpha_t$. The glass transition is not itself a phase transition, but a laboratory phenomenon, and depends on the cooling rate and the boundary conditions. Our calculations were obtained for systems containing 13824 atoms at a cooling rate of 10 K per ps from the liquid melt with 100 ps annealings every 100 K.

\begin{figure}[ht]
\centering
\subfigure[]{
 {\includegraphics[width=0.45\textwidth]{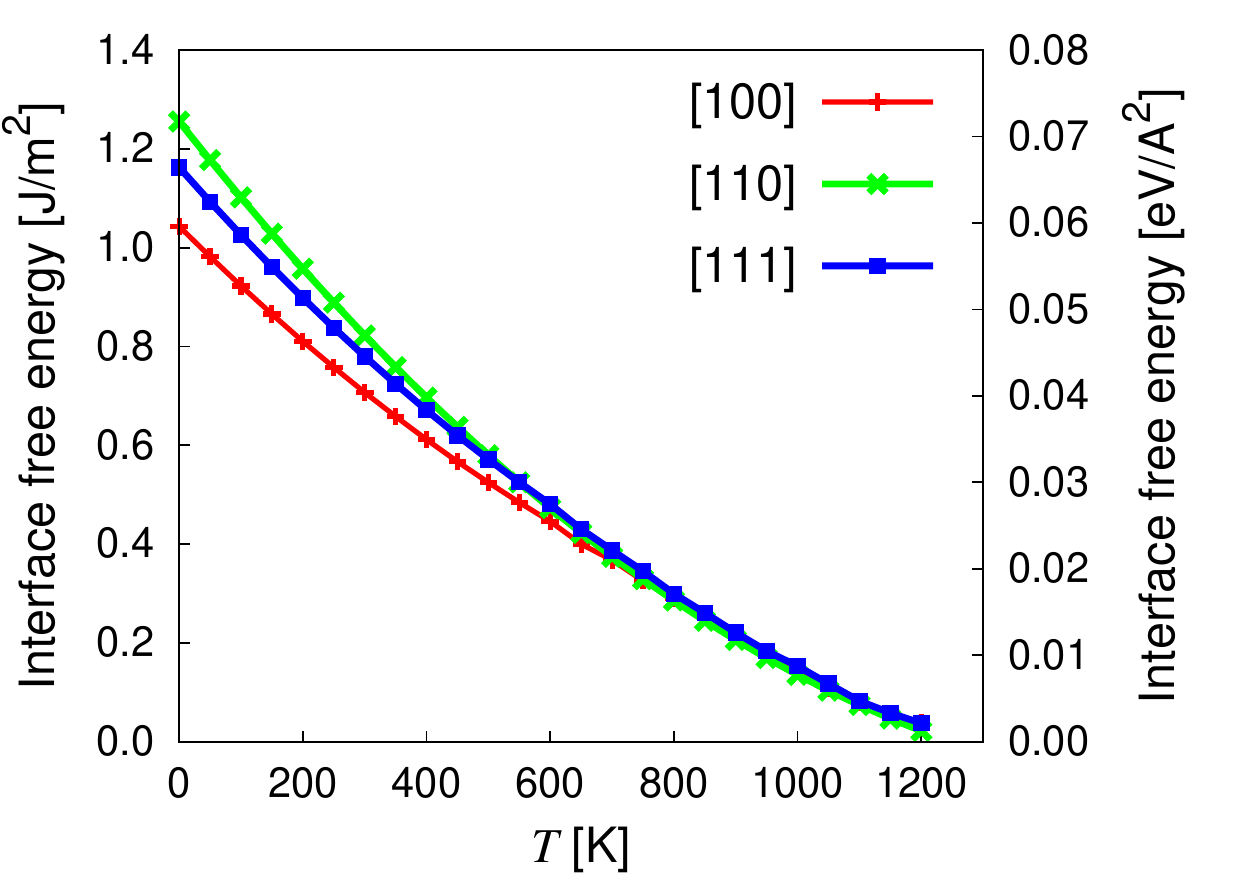}}
    \label{Fig:InterfaceEnergy_Temperature}
}
\subfigure[]{
  {\includegraphics[width=0.45\textwidth]{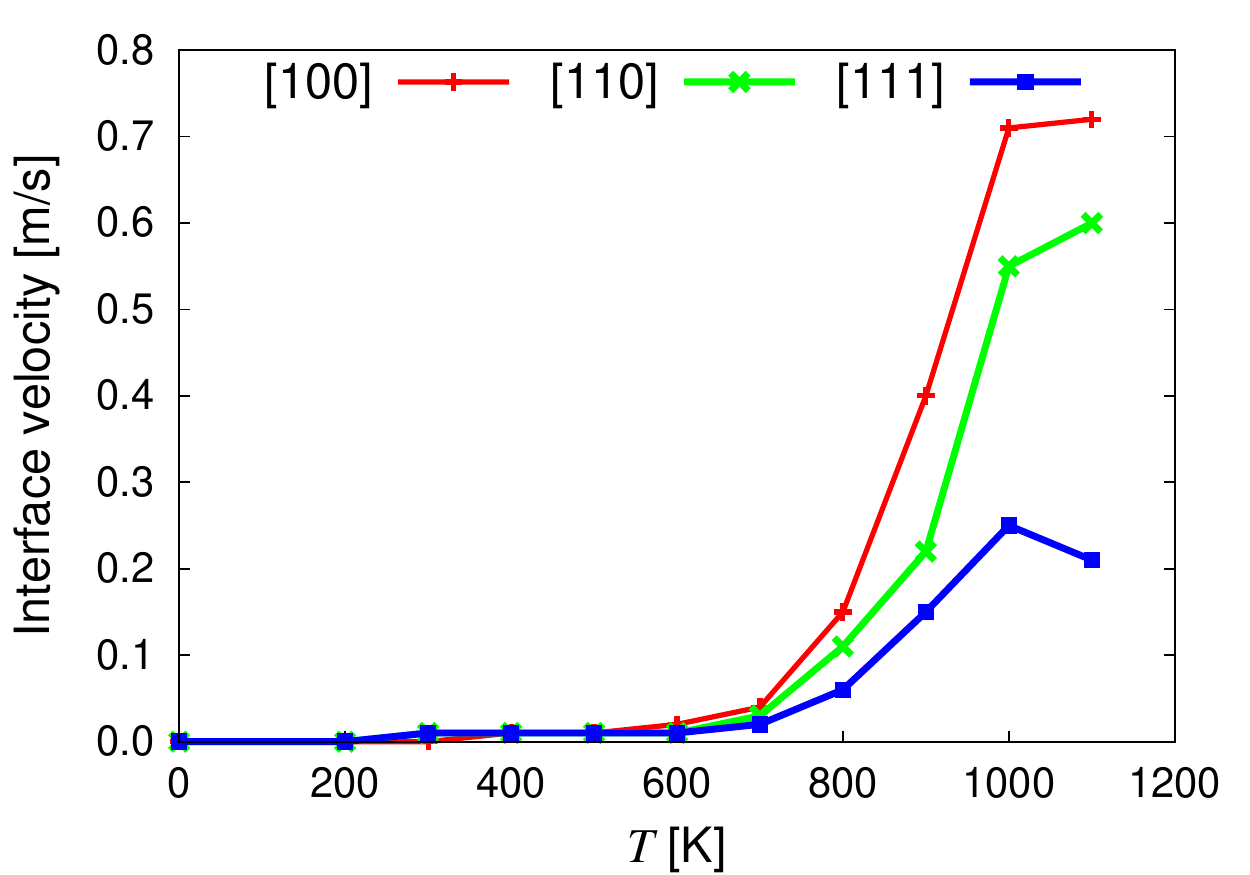}}
    \label{Fig:InterfaceVelocity_Temperature.pdf}
}
\caption{Calculations of  a planar interface at zero pressure: \subref{Fig:InterfaceEnergy_Temperature} surface free energy and \subref{Fig:InterfaceVelocity_Temperature.pdf} front velocity as a function of temperature for three surface normals.}
\label{Fig:InterfaceEnergy_InterfaceVelocity}
\end{figure}

Next, we construct a bilayer system consisting of crystalline and amorphous half crystals to measure interface properties. We calculate interfacial free energies and interface velocities as a function of temperature. To ascertain the orientation dependence of the crystal-amorphous interface, three different crystallographic directions have been considered, namely [100], [110] and the [111]. Figure \ref{Fig:InterfaceEnergy_Temperature} reveals the temperature evolution of the interfacial free energy, which is self-consistent in that it vanishes at the melting point. The $[100]$ orientation is the most stable one at low temperatures, followed by the $[111]$ and the $[110]$ directions. At the glass transition temperature, however, all interfacial free energies converge, likely due to a `wetting' effect of the crystalline surface in contact with a liquid-like system. For their part, the interface velocities also display a behavior qualitatively similar to the self-diffusion coefficient for intermediate temperatures (cf.~Fig.~\ref{fig:diff}). This amounts to practically zero mobility below $T_g$ and a monotonic increase above it. For very large temperatures though, the driving force decreases till vanishing at the melting temperature, and induces a decrease in the velocity. The compromise between driving force and mobility leads to a peak in the interface velocity around $1100$ K.

The last parameter to be computed from atomistic simulations is the energy barrier along the crystal-amorphous transformation path, $B(T)$, introduced in Section \ref{phase}. This calculation involves defining a suitable reaction coordinate between the initial --amorphous-- and final --crystalline-- states. A natural choice of initial and final states is to start from a crystalline specimen and melt it to the liquid state, followed by a sensible cooling into the amorphous structure. However, in most constrained minimization techniques used for this purpose --e.g.~nudged elastic band or drag methods--, the reaction coordinate vector is constructed as a linear interpolation of all degrees of freedom within each end state. Due to the sizable amount of mixing and thermal transport occurring during the melting and cooling phases, however, we have seen that this construction leads to large variations in the reaction coordinate that make convergence difficult (and the physical meaning of the procedure questionable). Instead, we proceed to use the bilayer geometry described above and let the interface propagate spontaneously turning the amorphous half system into a crystalline one. 
Once the conversion is complete, we extract a volumetric sample from within the converted region and use it as one of the end states, while using the corresponding atoms contained within that volume in the amorphous region as the second end state. Replicas for constrained minimization are generated by linear interpolation between both end states.
This procedure establishes a direct link between both structures while minimizing the amount of internal transport, providing a more physical transition pathway.

\begin{figure}[ht]
\centering
\includegraphics[width=0.48\textwidth]{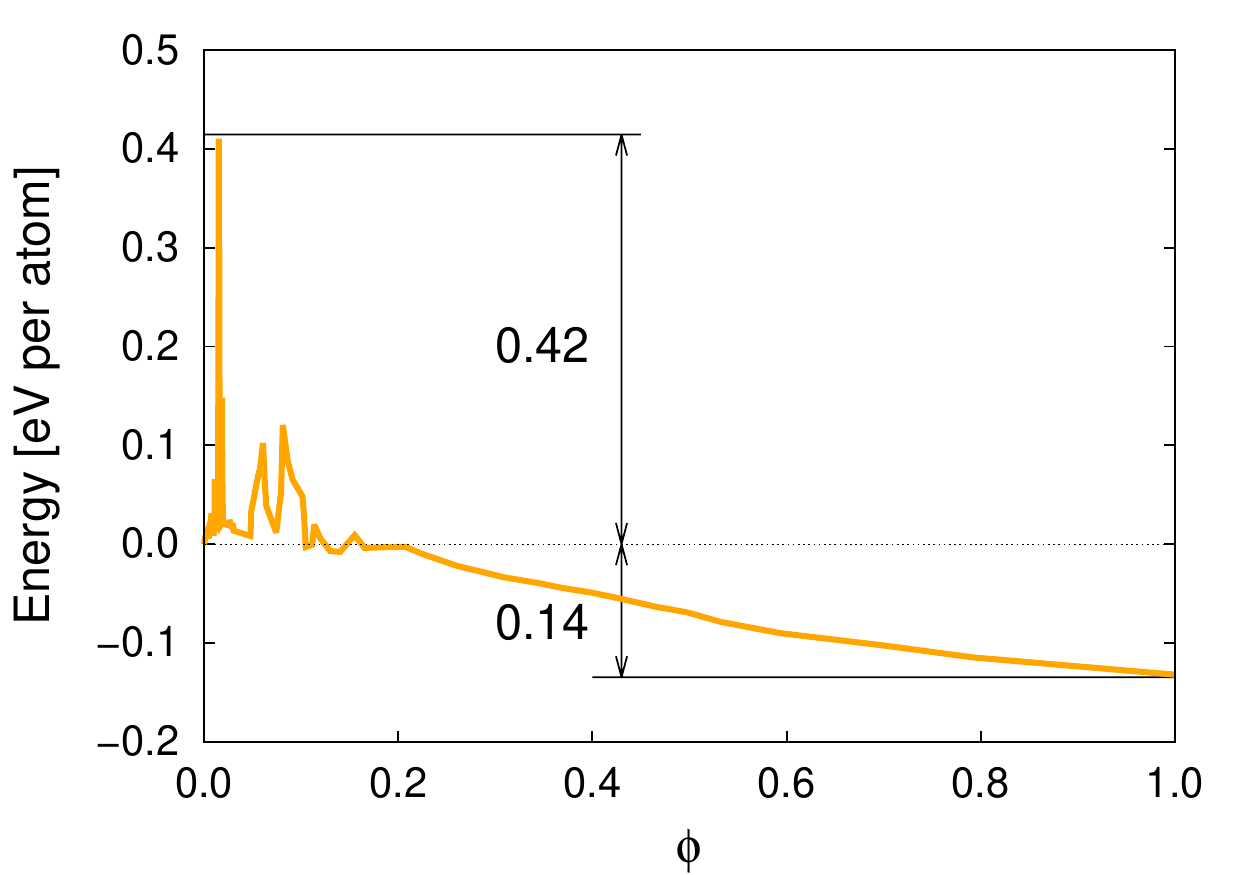}
\label{fig:bt}
\caption{Minimum energy path at zero temperature and pressure for conversion of an amorphous system into an ordered crystal. A barrier of 0.49 eV per atom is seen to separate both phases, while the energy difference between the stable amorphous and crystalline is 0.14 eV per atom, consistent with Fig.\ \ref{Fig:FreeEnergy_Temperature} at 0 K. The roughness in the energy path is a consequence of the multidimensional nature of the transition, with many atoms undergoing transformations at once.}
\end{figure}

Figure \ref{fig:bt} shows the relaxed minimum energy path at 0 K for a system containing 13824 atoms and 16 replicas. The static energy barrier is  approximately $B(0)\approx0.49$ eV per atom. Finding the temperature dependence intrinsic in $B(T)$ is not trivial and requires using elaborate thermodynamic integration techniques ({\it e.g.} refs.\ \cite{carter1989,gilbert2013}). Here, for lack of a better estimate, we assume that $B(T)=B(0)$.

\subsection{Numerical analyses of the phase field model}

The evolution equation for the phase field variable, c.f.~Eq.~(\ref{Eq:phi_constantT}), is numerically solved in this work by means of a semi-implicit, backward Euler discretization scheme in time
\begin{equation}
\alpha \epsilon^2 \frac{\phi^{j+1}-\phi^j}{\Delta t} = \epsilon^2\ \nabla^2 \phi^{j+1} - g'_{dw}(\phi^j)+ \Delta g\ q'(\phi^j),
\end{equation}
where $\Delta t$ is the time step. The resulting ordinary differential equation is then solved in space via a first order finite difference scheme followed by a discrete Fourier transform (\verb+fftw+ package \citep{FFTW05}). The grid size is denoted by $h$.

In this section we illustrate the convergence of the numerical scheme to known analytical solutions and demonstrate good agreement with the theoretical spatial and temporal convergence rates.

\subsubsection{Static one-dimensional solution at constant temperature}
We first consider a planar front separating a two phase region at constant temperature with an energy landscape described by a perfect double-well potential, i.e.~with $\Delta g = 0$. The equilibrium solution of this system is characterized by Eq.~(\ref{Eq:phi_constantT}), that is, 
\begin{equation}
\alpha \epsilon^2 \frac{\partial \phi}{\partial t} = \epsilon^2 \nabla^2 \phi - 32 \phi(1- \phi)(1-2 \phi).
\end{equation}

This equation is analytically soluble and gives a static interface with the following profile
\begin{equation}\label{Eq:Analytic_phi}
\phi(x) = \frac{1}{2} - \frac{1}{2} \text{tanh} \left(\frac{x}{\epsilon} \sqrt{8}\right),
\end{equation}
where the $x$ axis is centered at the interface and orthogonal to it. For this solution, $\epsilon$ exactly corresponds to the interface thickness, measured as the support of $0.056 < \phi < 0.944$.

\begin{figure}
\begin{center}
    {\includegraphics[width=0.45\textwidth]{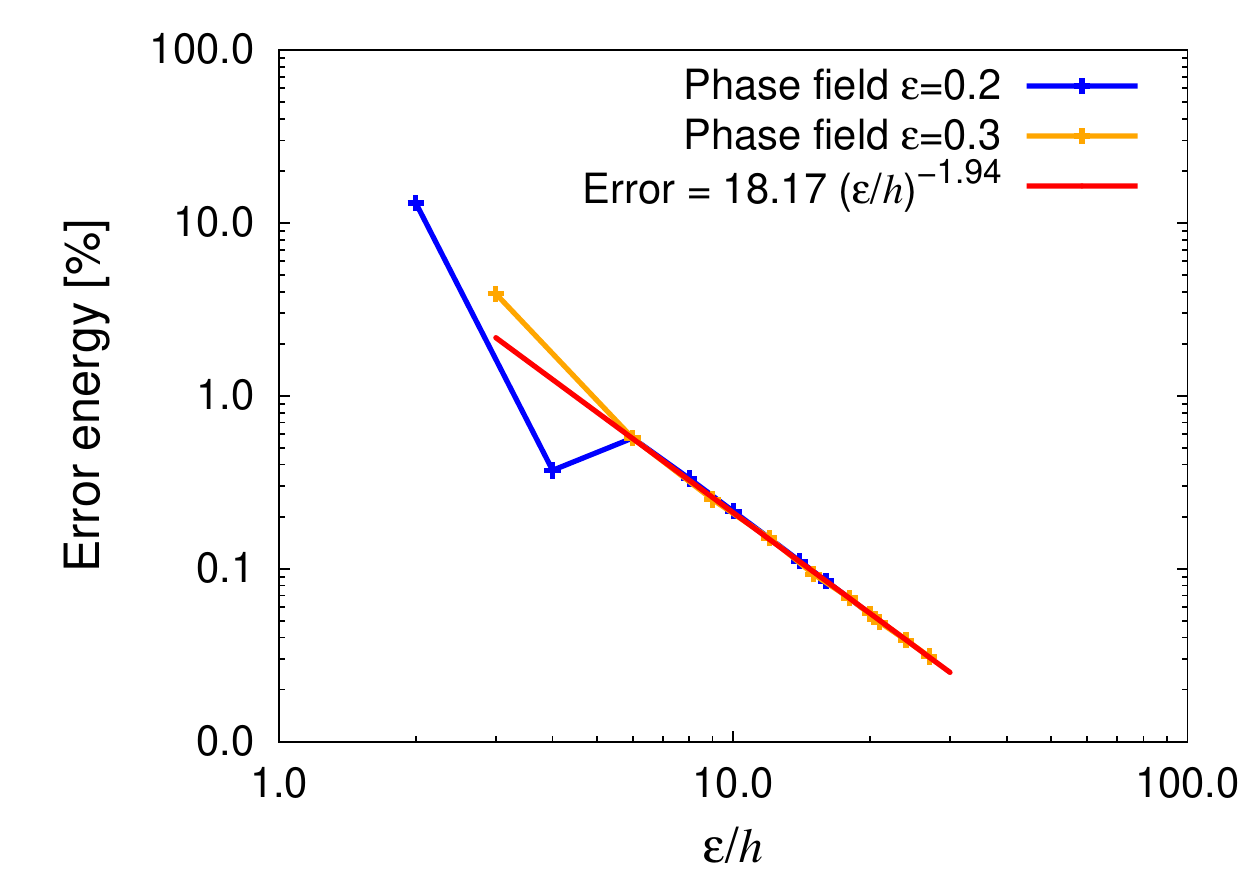}}
    \caption[]{Energy error as a function of the normalized mesh size. Parameters of the simulation: $\epsilon = 0.2$ and $0.3$, $\alpha = 1$, $\Delta t = 0.001$.}
    \label{Fig:1D_DoubleWell_EnergyError}
\end{center}
\end{figure}

Figure \ref{Fig:1D_DoubleWell_EnergyError} shows the error with respect to the analytic solution with the following energy measure

\begin{equation}
\begin{split}
&\text{Error } = \frac{\text{Energy}_{\text{analytic}}-\text{Energy}_{\text{PhaseField}}}{\text{Energy}_{\text{analytic}}} 100 \\
&\text{Energy}_{\text{analytic}} = \int_{x:\phi = 0.95}^{x:\phi = 0.05} 16 \phi^2(1-\phi)^2 \, dx.
\end{split}
\end{equation}

We observe a quadratic convergence of the energy as it is expected from the first order finite difference scheme used in these simulations.

\subsubsection{Steady state one-dimensional solution at constant temperature}

\begin{figure}
\begin{center}
    {\includegraphics[width=0.45\textwidth]{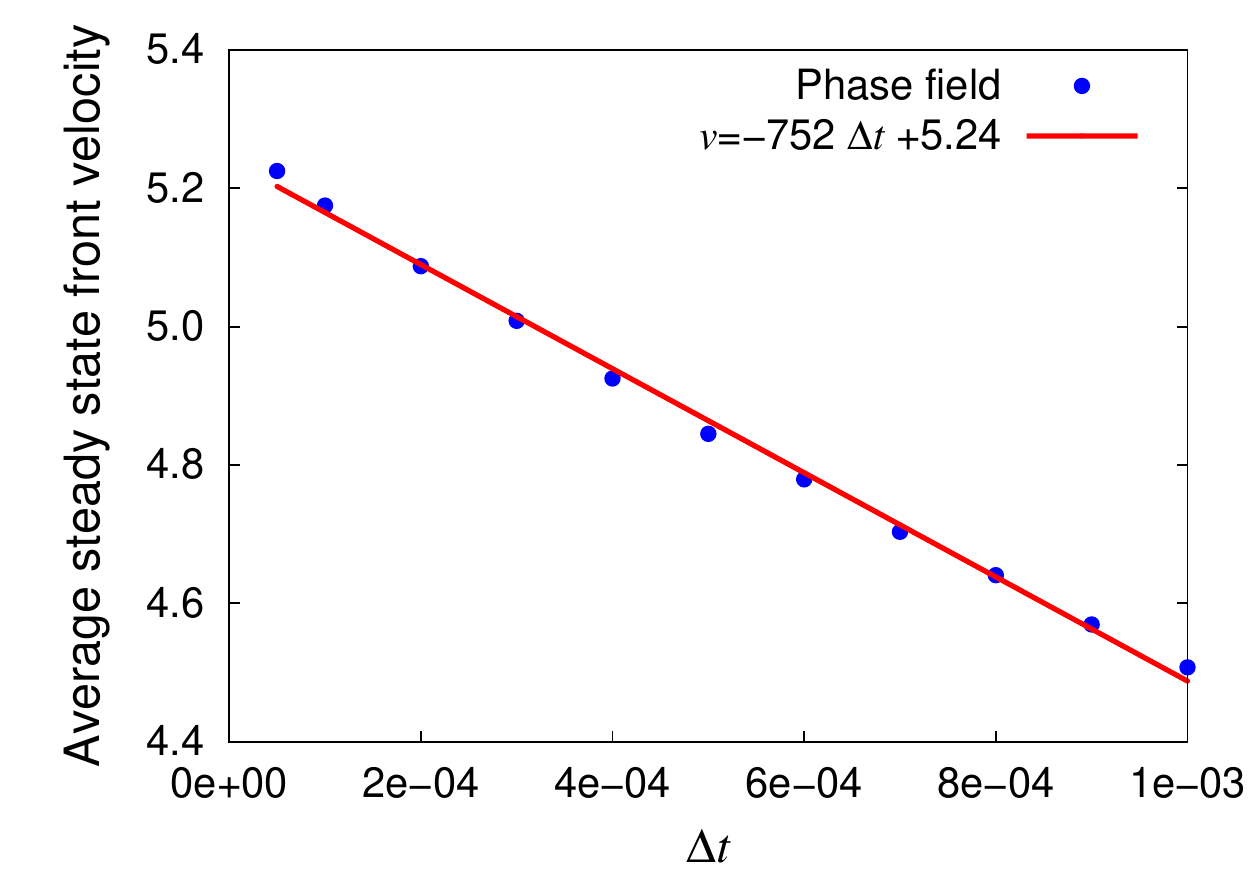}}
    \caption[]{Average steady state velocity with respect to time step. $\epsilon = 0.2$, $\alpha = 1$, $\Delta \bar{g} = 1$, $h=0.01$}
    \label{Fig:1D_InclinedDoubleWell_FrontVelocity_DT}
\end{center}
\end{figure}

\begin{figure}[ht]
\centering
\subfigure[Average velocity of the steady state numerical solution with respect to the mesh size for a constant time step.]{
 {\includegraphics[width=0.45\textwidth]{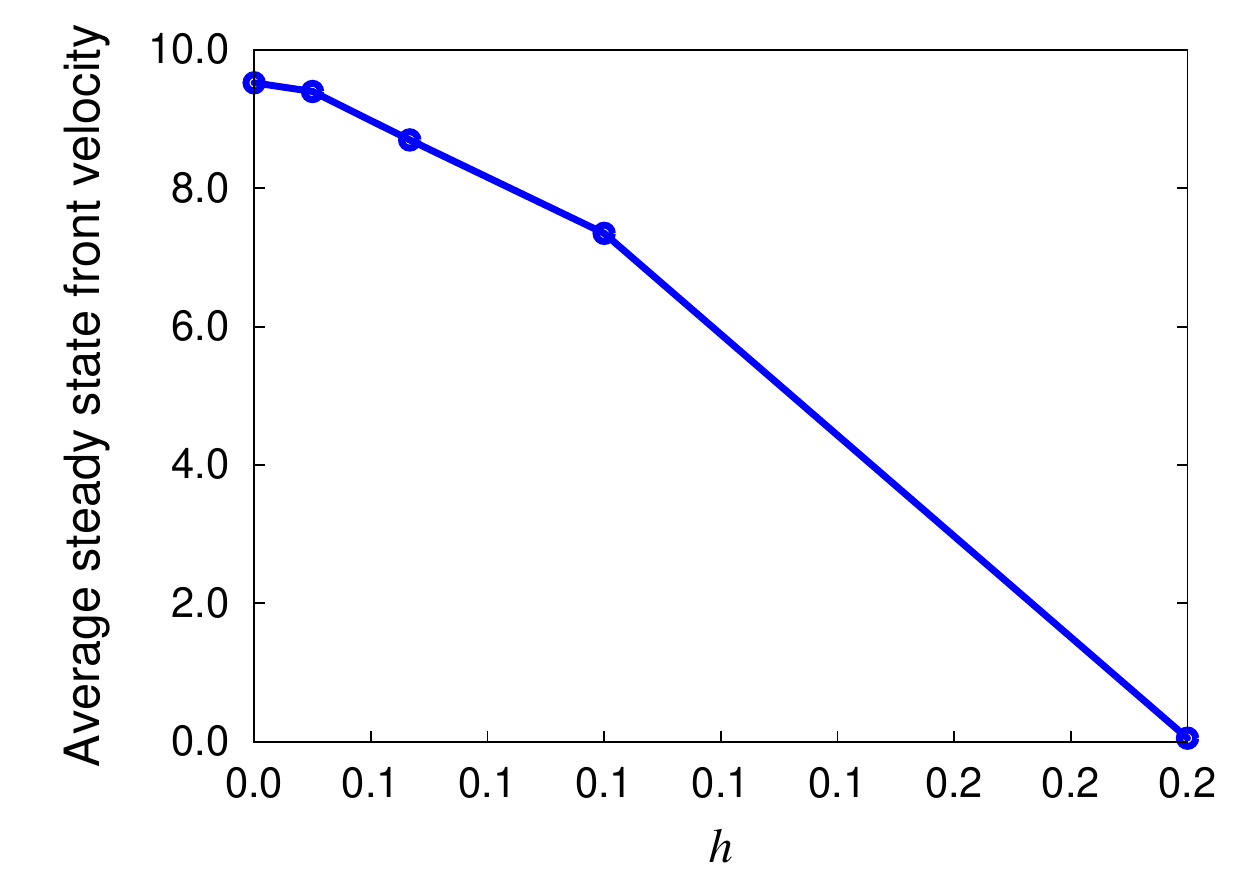}}
    \label{Fig:1D_InclinedDoubleWell_FrontVelocity_NE}
}
\subfigure[Evolution of the front velocity with time for several mesh sizes.]{
 {\includegraphics[width=0.45\textwidth]{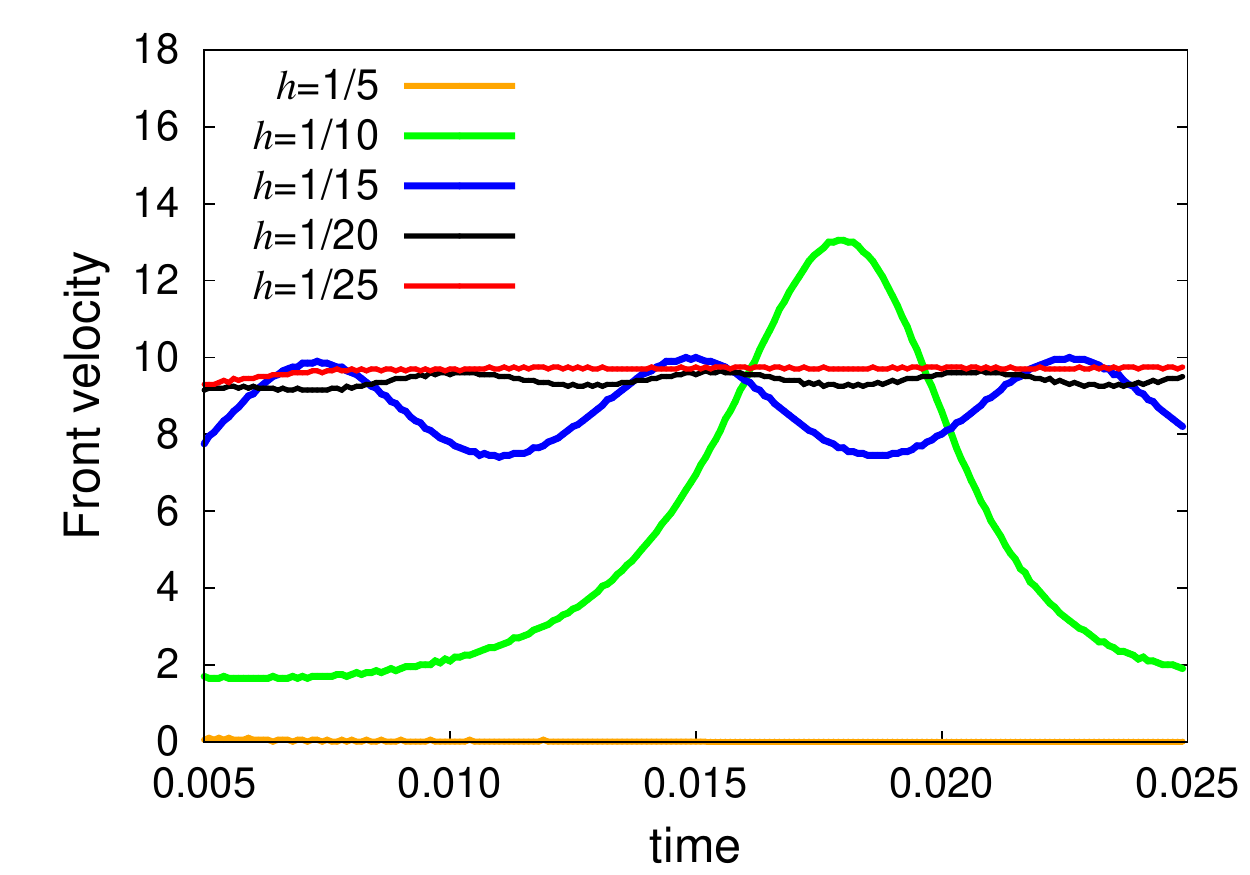}}
    \label{Fig:1D_InclinedDoubleWell_FrontVelocity_MultipleNE}
}
\caption{Numerical behavior of a planar amorphous-crystalline front with increasing mesh size. Parameters of the simulation:   $\epsilon = 0.2$, $\alpha = 1$, $\Delta t = 0.0001$, $\Delta \bar{g} = 2$. }
\label{Fig:1D_InclinedDoubleWell_FrontVelocity_MultipleNE_a}
\end{figure}

A difference in free energy between the two phases ($\Delta g> 0$) will induce a planar interface to move until a constant velocity is reached, as discussed in Sec.~\ref{Sec:Methods}.  Figure \ref{Fig:1D_InclinedDoubleWell_FrontVelocity_DT}  shows the value of the numerical steady state velocity as a function of the time step for a constant (small) mesh size.  As expected from a first order time scheme, the average velocity converges linearly with the time step. However, if the mesh size used for the spacial discretization is too large, the numerical solution does not lead to a steady state velocity. Rather, the underlying mesh acts as a periodic potential inducing oscillations in the steady state numerical solution, c.f.~Fig.~\ref{Fig:1D_InclinedDoubleWell_FrontVelocity_NE}. Additionally, the mean velocity decreases with the mesh size for a given constant time step, and can even vanish for very large mesh sizes, c.f.~Fig.~\ref{Fig:1D_InclinedDoubleWell_FrontVelocity_MultipleNE}. This behavior, previously reported by other authors \citep{KarmaRappel1998}, is characteristic of the scheme used for solving the ordinary differential equation at a given constant time step. Equivalent simulations using finite elements indicate that the average steady state velocity may increase with the mesh size for sufficiently large driving forces.

\subsection{Parameters of the phase field model}

\begin{figure}[ht]
\centering
\subfigure[]{
 {\includegraphics[width=0.45\textwidth]{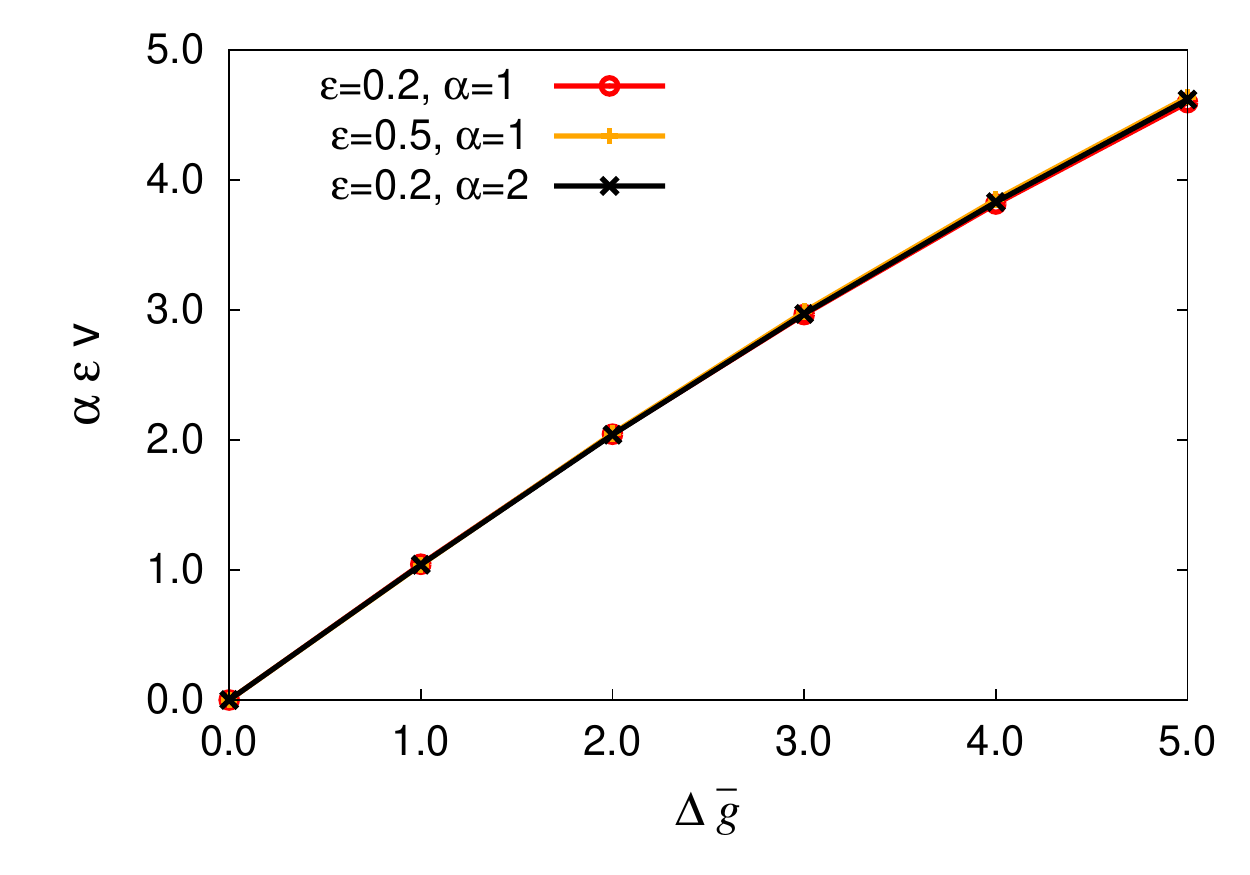}}
    \label{Fig:InvarianceRelation_Velocity}
}
\subfigure[]{
 {\includegraphics[width=0.45\textwidth]{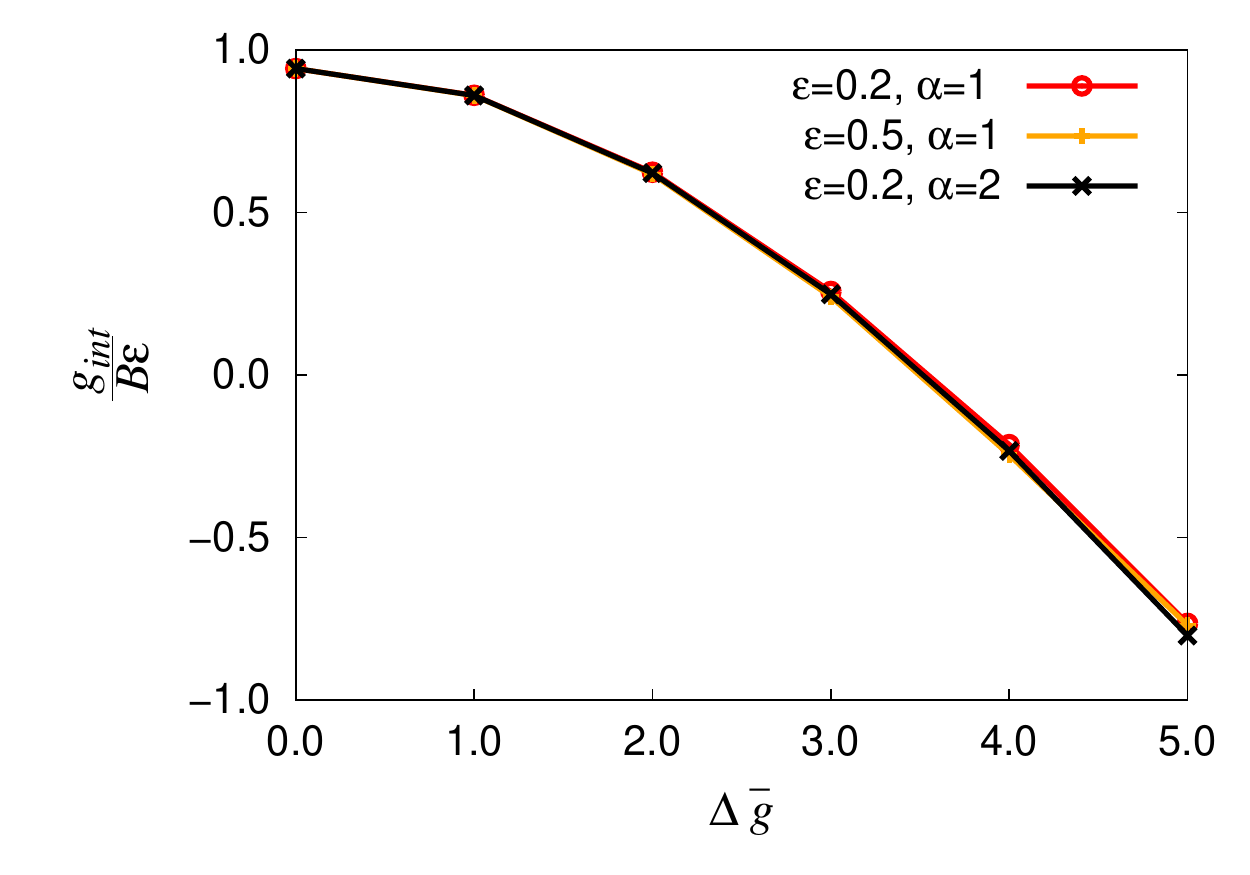}}
    \label{Fig:InvarianceRelation_InterfaceEnergy}
}
\caption{Invariance relations for the steady state solution of a planar crystalline-amorphous interface:   \subref{Fig:InvarianceRelation_Velocity} non-dimensional interface velocity and \subref{Fig:InvarianceRelation_InterfaceEnergy} free energy versus non-dimensional driving force. Parameters of the simulations: $\Delta t = 5\times10^{-5}$, $h = 0.005$.}
\end{figure}

The functions $I_v(\Delta \bar{g})$ and $I_g(\Delta \bar{g})$ described in Eqs.~(\ref{Eq:v_int}) and (\ref{Eq:f_int}), can now be computed with a controlled numerical error lower than 1$\%$. In Figures \ref{Fig:InvarianceRelation_Velocity} and \ref{Fig:InvarianceRelation_InterfaceEnergy} we plot each invariant for different values of the phase field parameters. For small values of $\Delta \bar{g}$, $I_v(\Delta \bar{g})$ and $I_g(\Delta \bar{g})$ can be approximated by
\begin{align} \label{Eq:FirstInvarance}
& \alpha \epsilon v \simeq \frac{\Delta g}{B},\\  \label{Eq:SecondInvarance}
&g_{int} \simeq \epsilon B.
\end{align}
The linear dependence given in Eq.~(\ref{Eq:FirstInvarance}) between the interface velocity and the free energy difference is often assumed in the literature \citep{song2012}. For its part, Eq.~(\ref{Eq:SecondInvarance}) is the result of the equipartition of the interface energy at zero velocity, as shown in Appendix \ref{Sec:Appendix_A}.

\begin{figure}[ht]
\centering
\subfigure[]{
 {\includegraphics[width=0.35\textwidth]{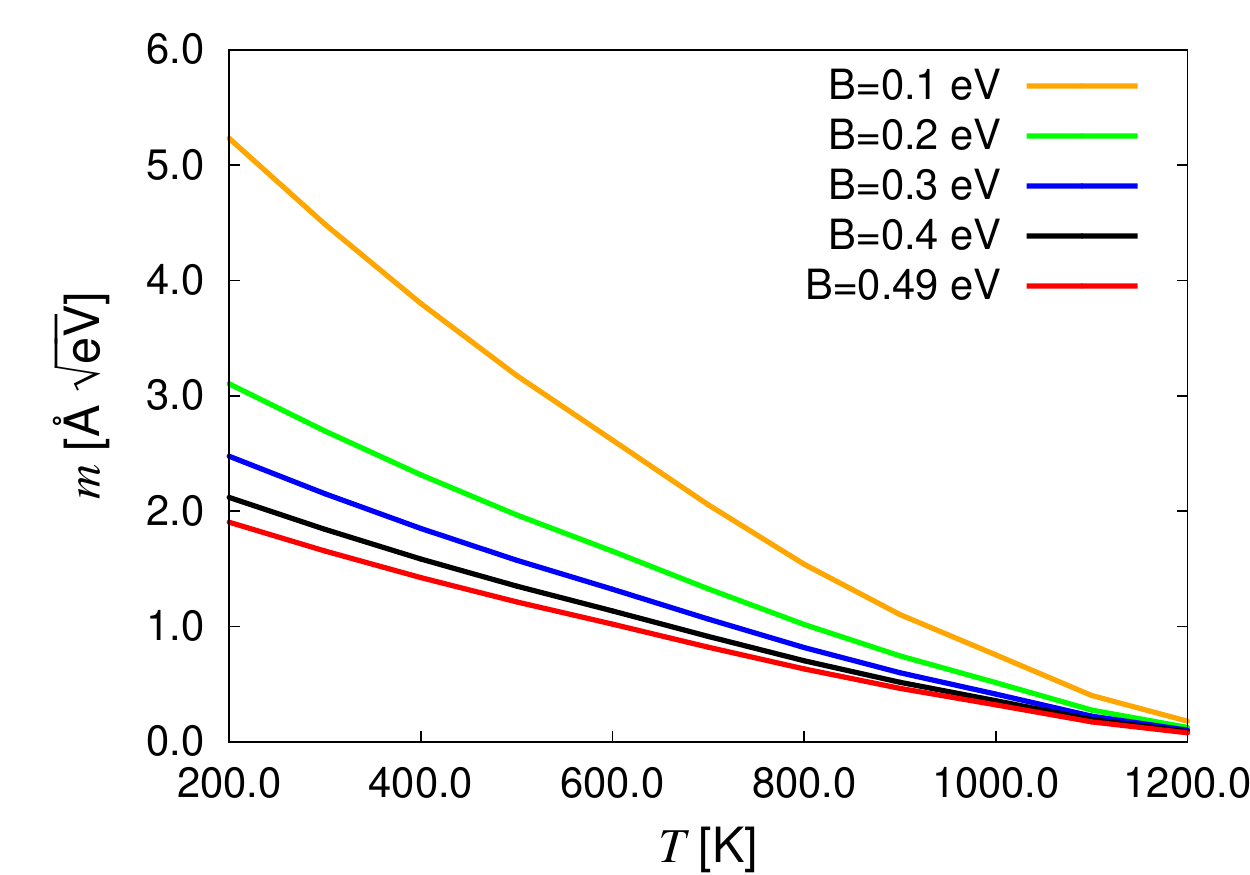}}
    \label{Fig:m_Temperature_111}
}
\subfigure[]{
  {\includegraphics[width=0.35\textwidth]{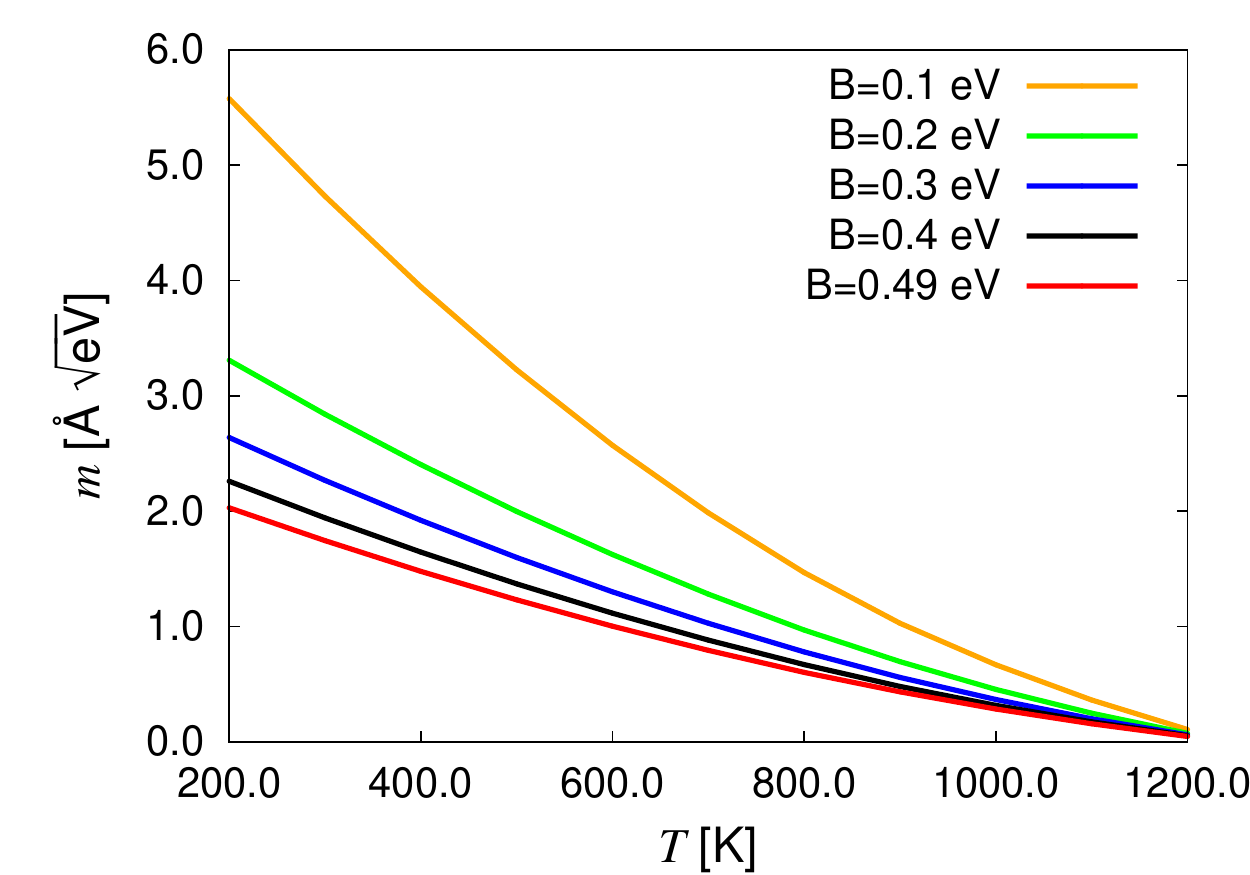}}
    \label{Fig:m_Temperature_110}
}
\subfigure[]{
 {\includegraphics[width=0.35\textwidth]{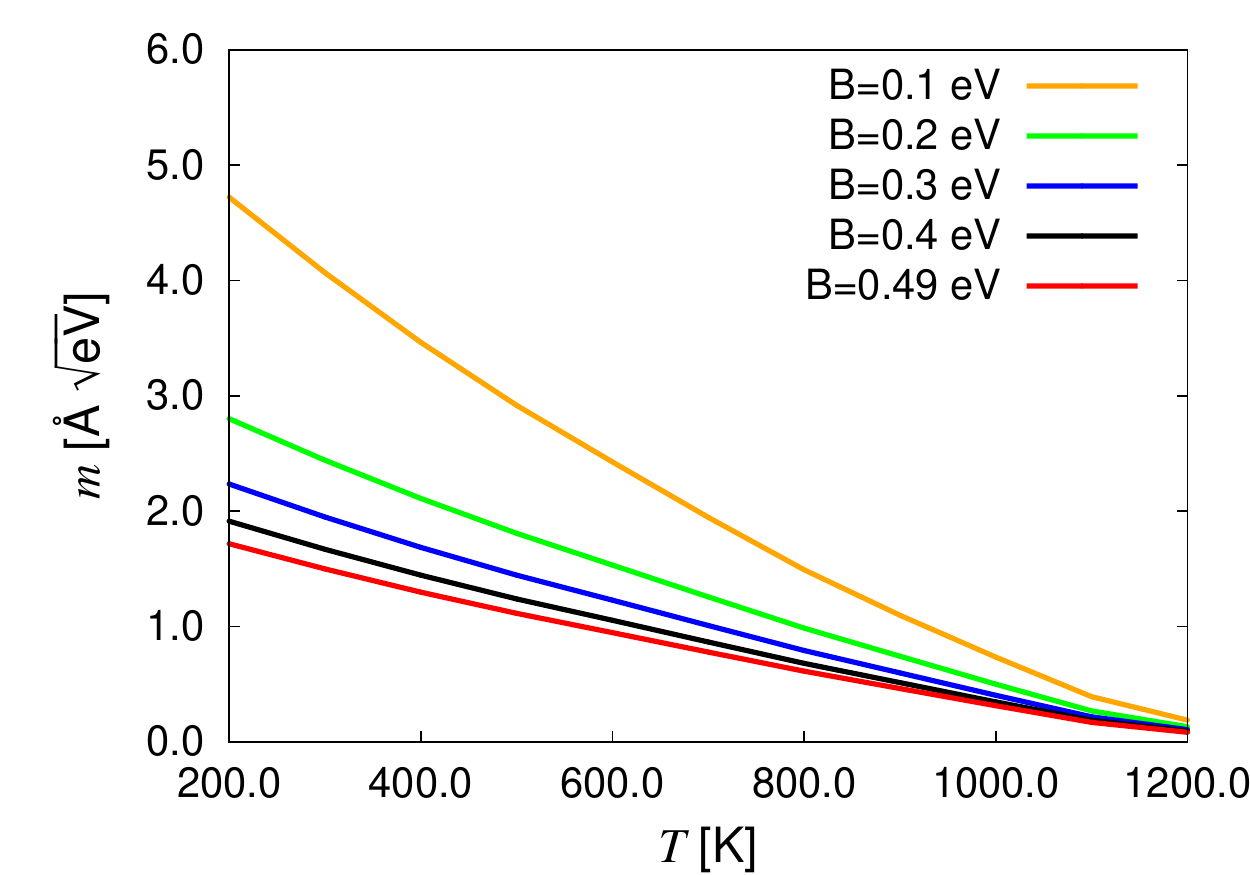}}
    \label{Fig:m_Temperature_100}
}
\caption{Phase field parameter $m$ as a function of temperature for several values of the free energy barrier $B$: \subref{Fig:m_Temperature_111} (111) surface, \subref{Fig:m_Temperature_110} (110) surface and \subref{Fig:m_Temperature_100} (100) surface.}
\label{Fig:m_Temperature}
\end{figure}

\begin{figure}[ht]
\centering
\subfigure[]{
 {\includegraphics[width=0.35\textwidth]{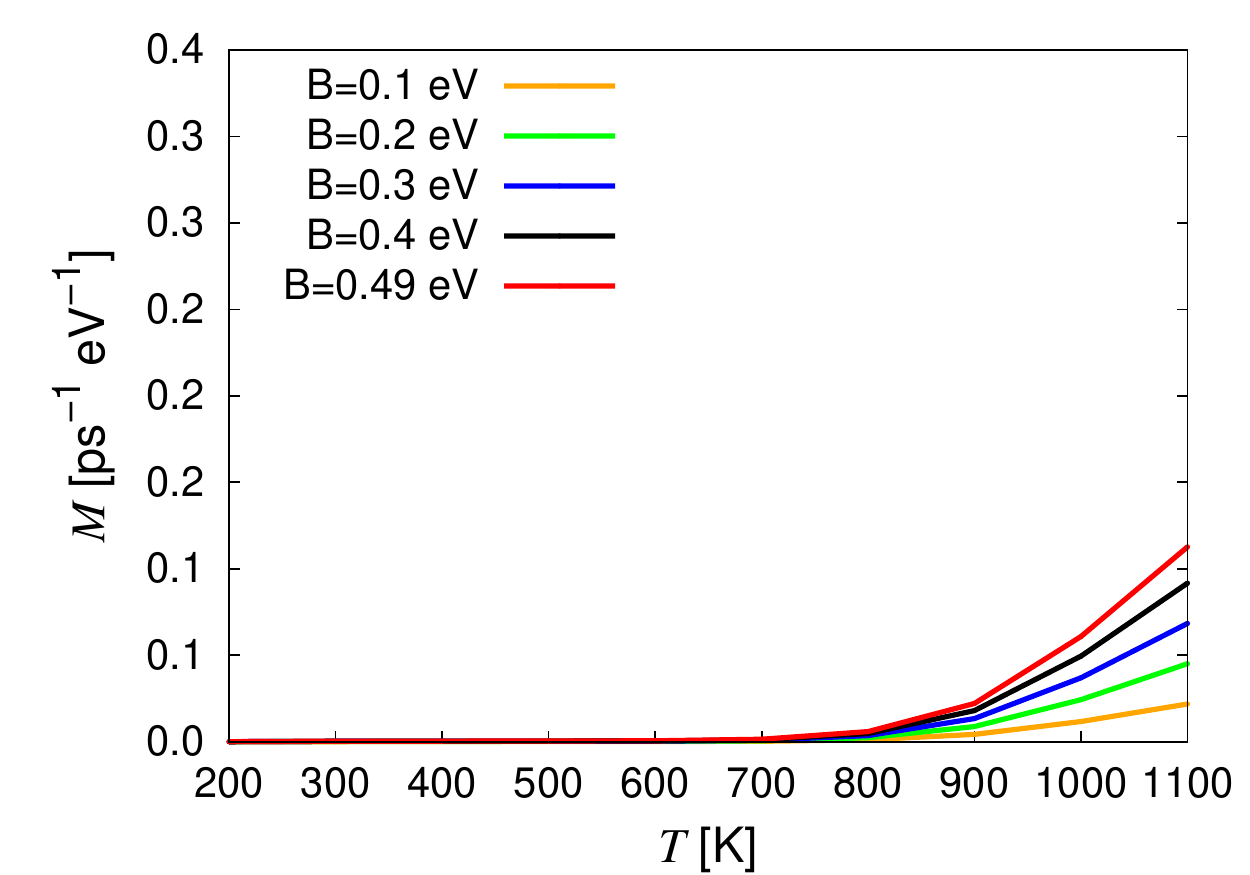}}
    \label{Fig:Mobility_Temperature_111}
}
\subfigure[]{
  {\includegraphics[width=0.35\textwidth]{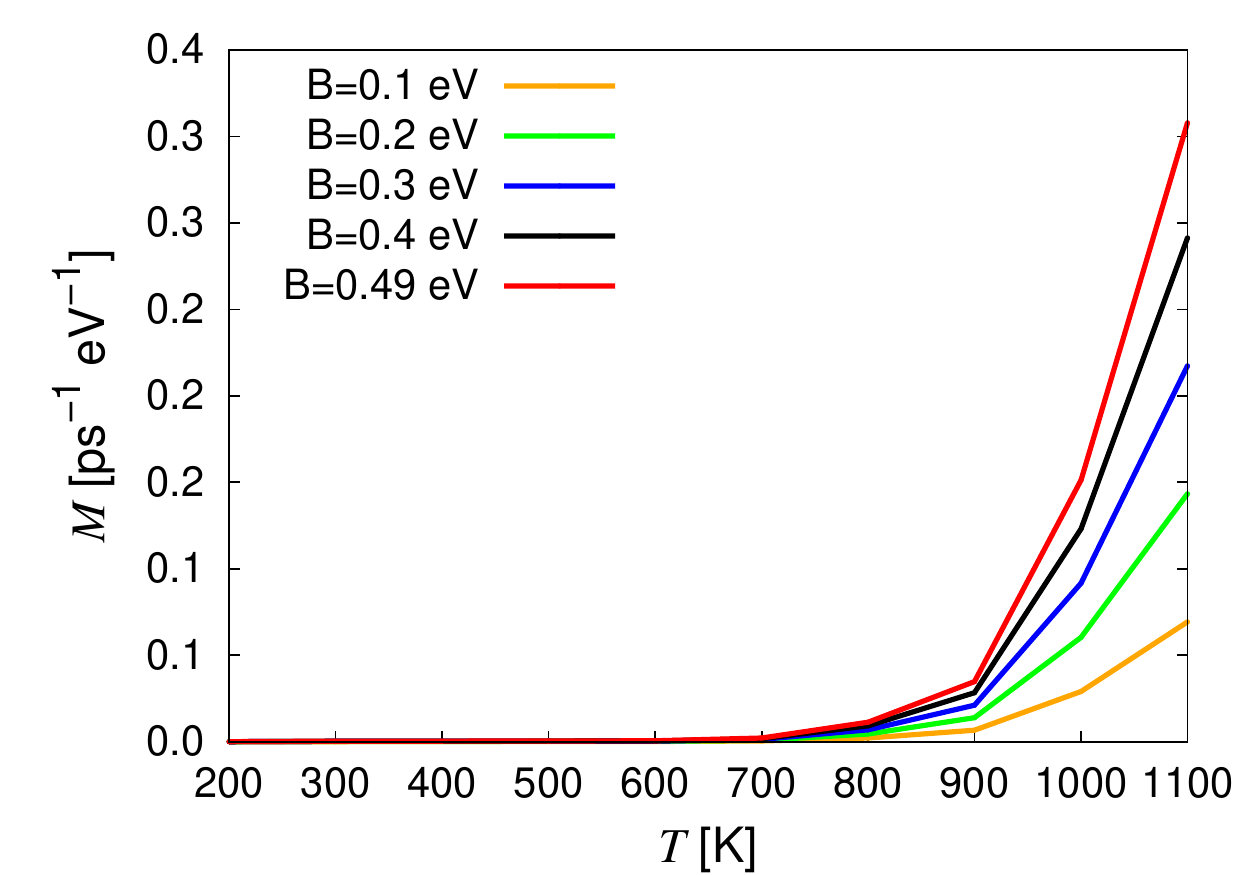}}
    \label{Fig:Mobility_Temperature_110}
}
\subfigure[]{
 {\includegraphics[width=0.35\textwidth]{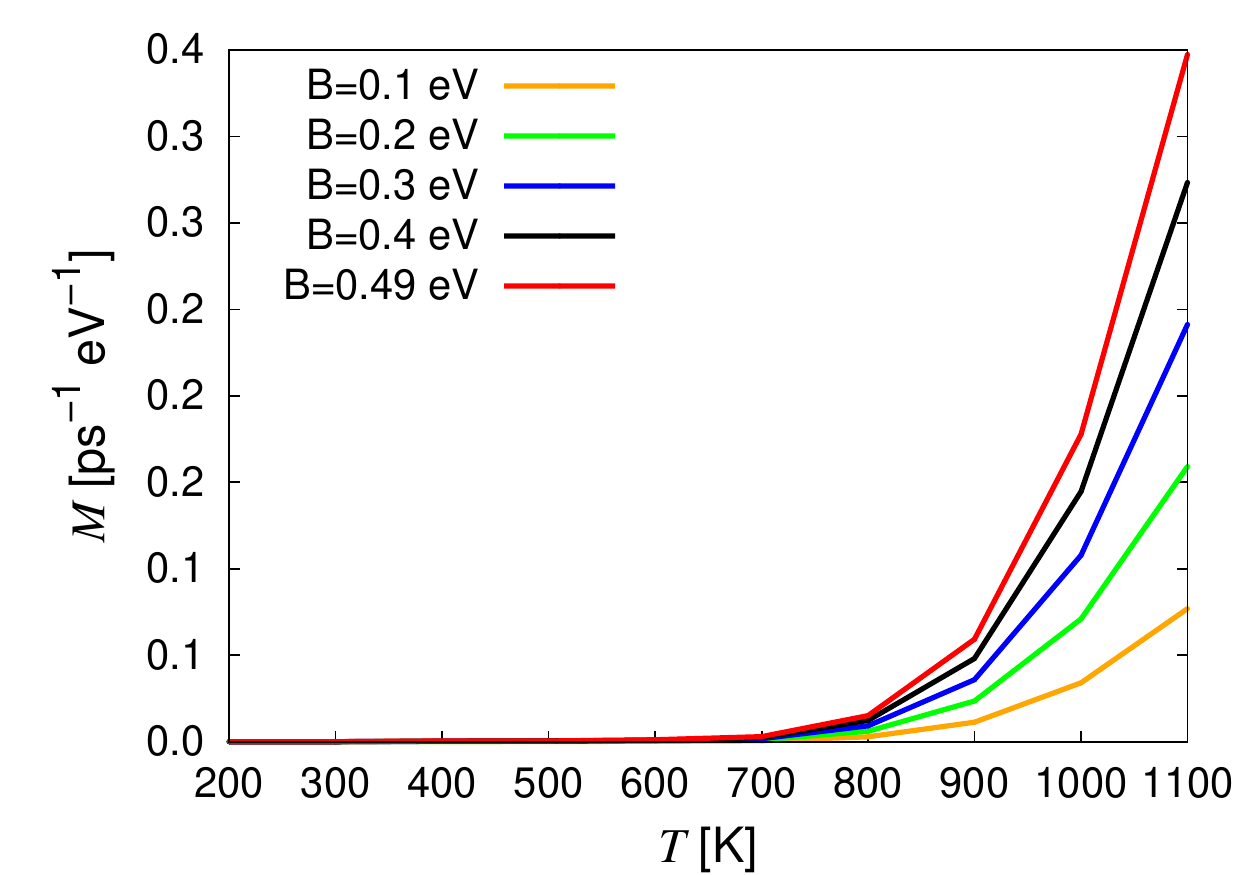}}
    \label{Fig:Mobility_Temperature_100}
}
\caption{Interface mobility as a function of temperature for several values of the free energy barrier $B$: \subref{Fig:Mobility_Temperature_111} (111) surface, \subref{Fig:Mobility_Temperature_110} (110) surface and \subref{Fig:Mobility_Temperature_100} (100) surface.}
\label{Fig:Mobility_Temperature}
\end{figure}

\begin{figure}[ht]
\centering
\subfigure[]{
 {\includegraphics[width=0.35\textwidth]{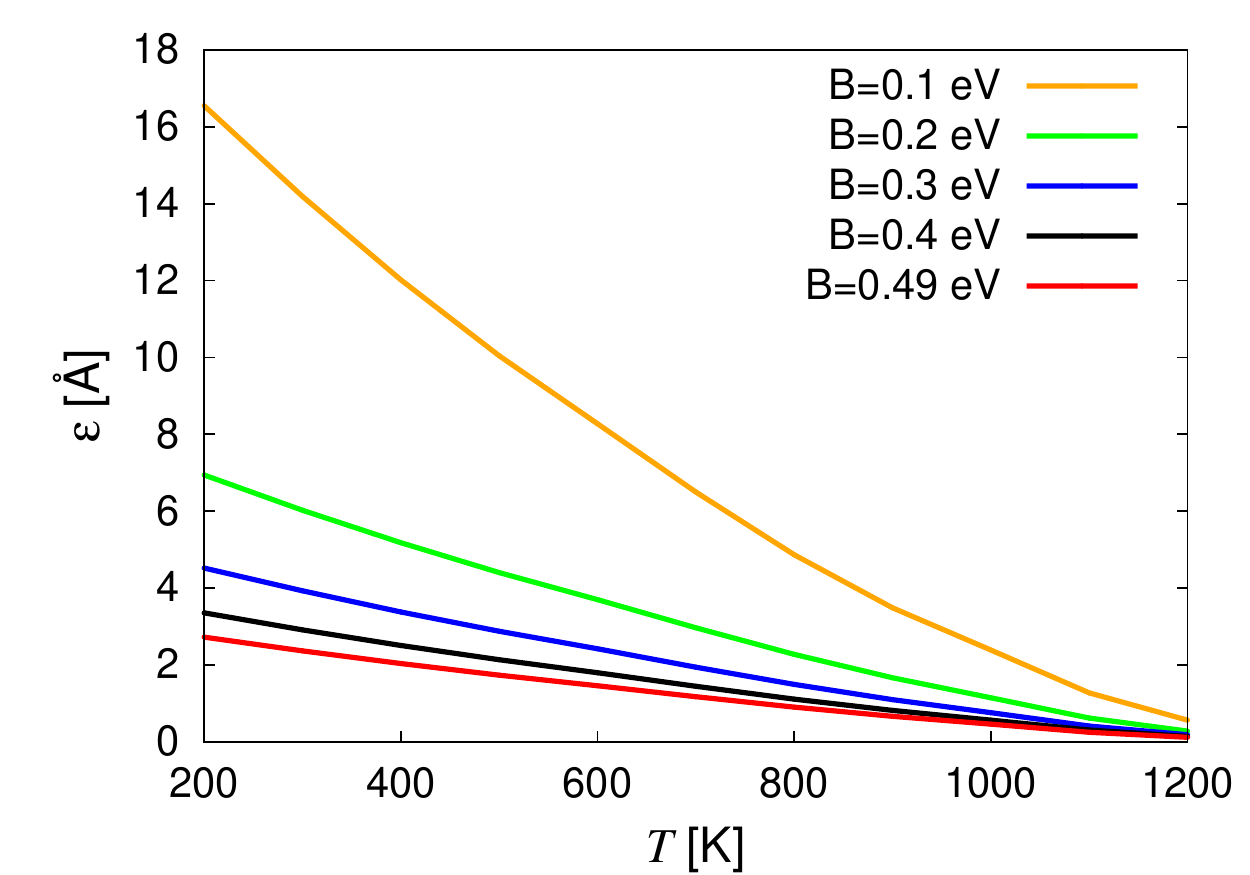}}
    \label{Fig:epsilon_temperature_111}
}
\subfigure[]{
  {\includegraphics[width=0.35\textwidth]{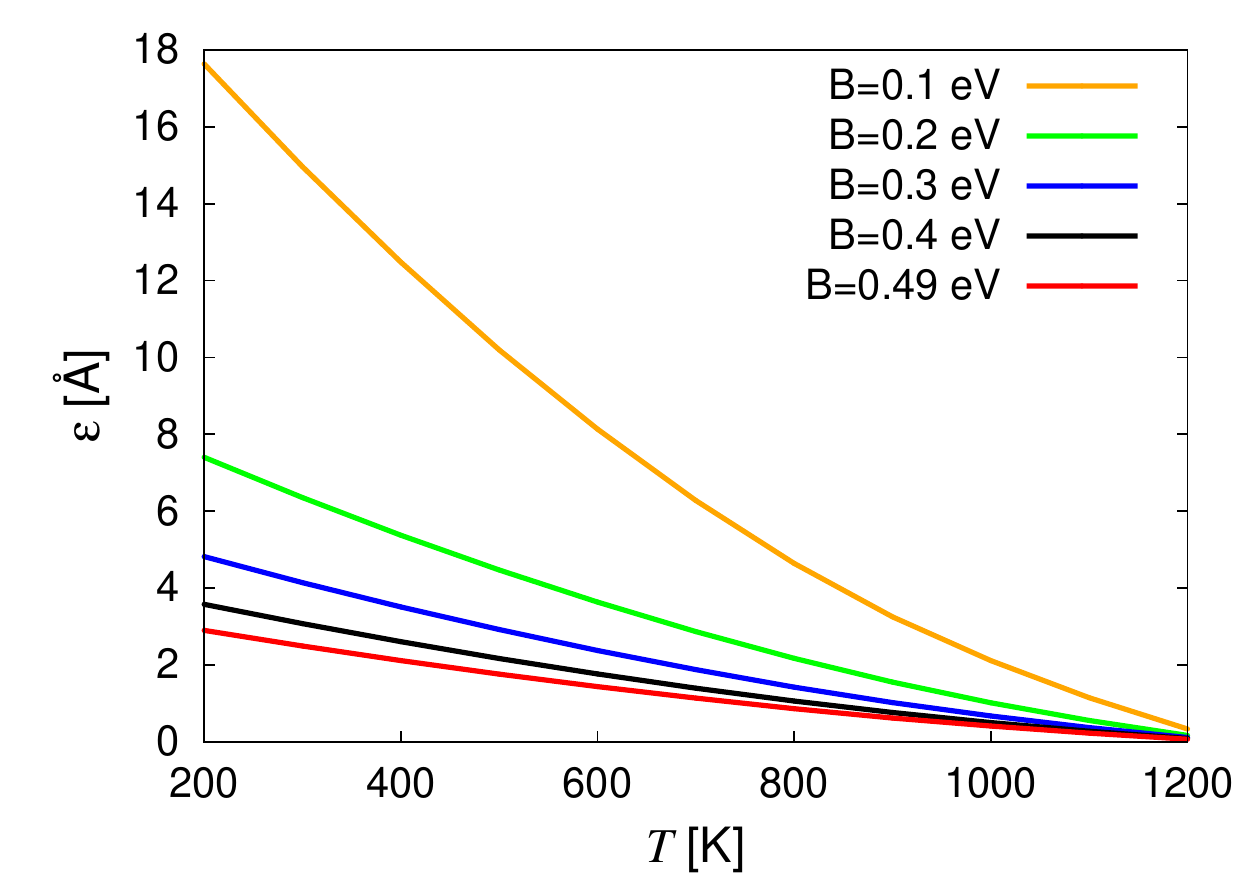}}
    \label{Fig:epsilon_temperature_110}
}
\subfigure[]{
 {\includegraphics[width=0.35\textwidth]{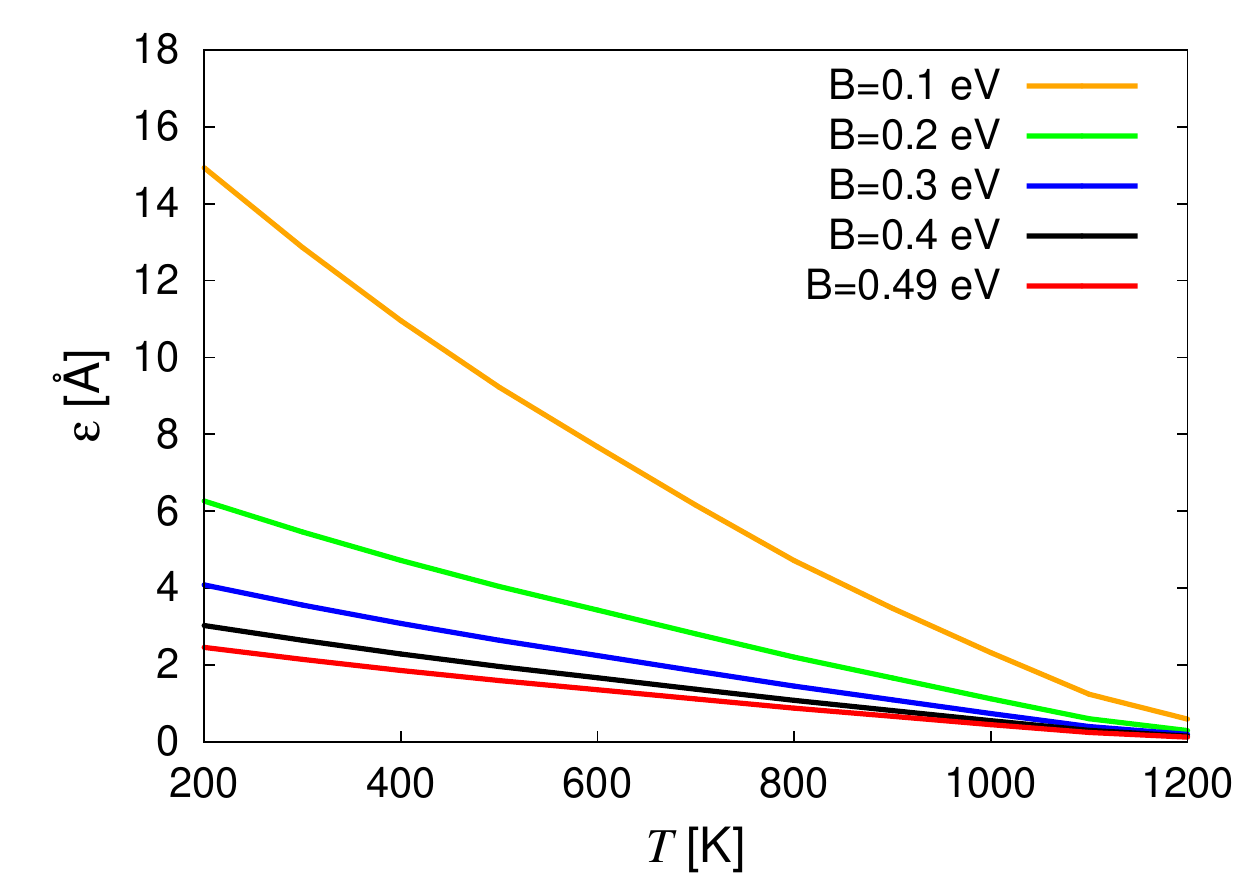}}
    \label{Fig:epsilon_temperature_100}
}
\caption{Phase field parameter $\epsilon$ as a function of temperature for several values of the free energy barrier $B$: \subref{Fig:epsilon_temperature_111} (111) surface, \subref{Fig:epsilon_temperature_110} (110) surface and \subref{Fig:epsilon_temperature_100} (100) surface.}
\label{Fig:epsilon_temperature}
\end{figure}

\begin{figure}[ht]
\centering
\subfigure[]{
 {\includegraphics[width=0.35\textwidth]{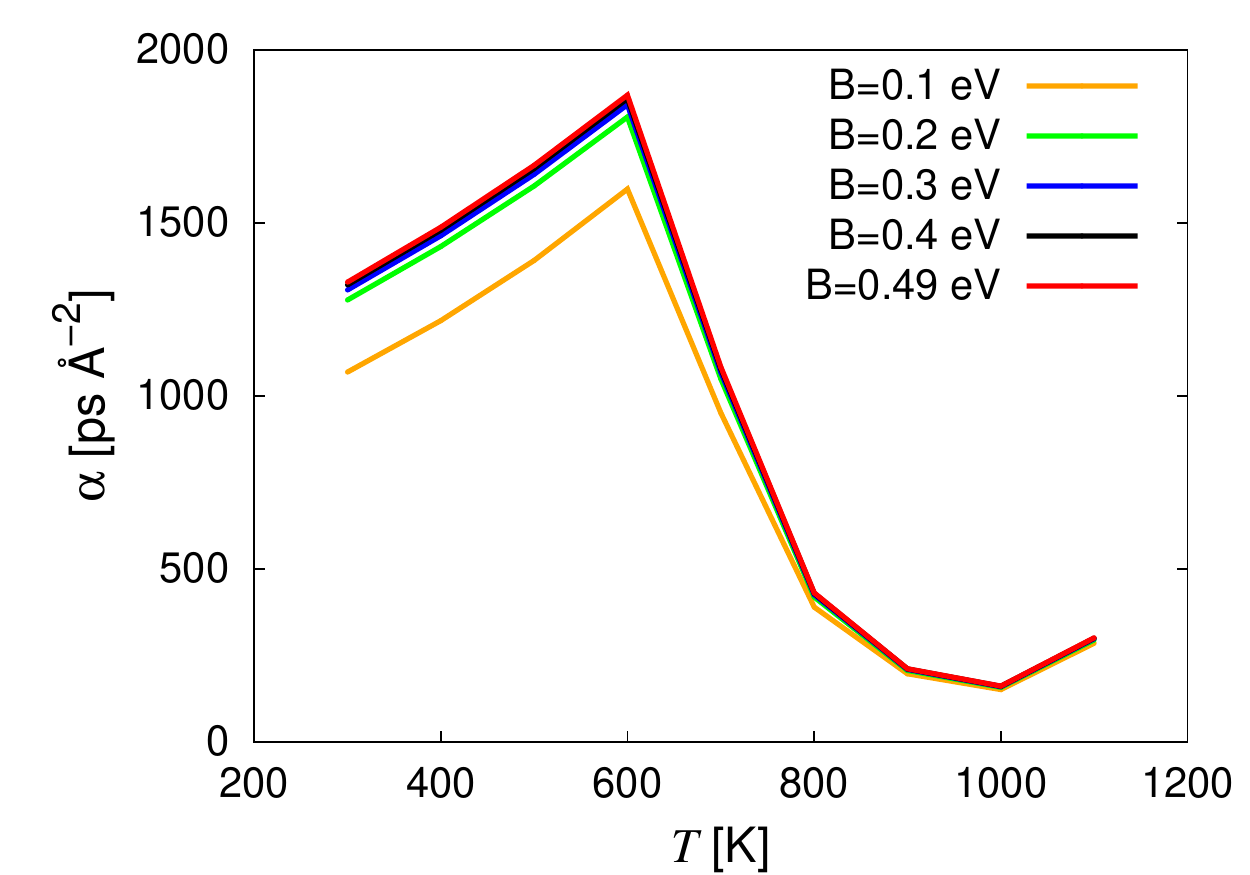}}
    \label{Fig:alpha_temperature_111}
}
\subfigure[]{
  {\includegraphics[width=0.35\textwidth]{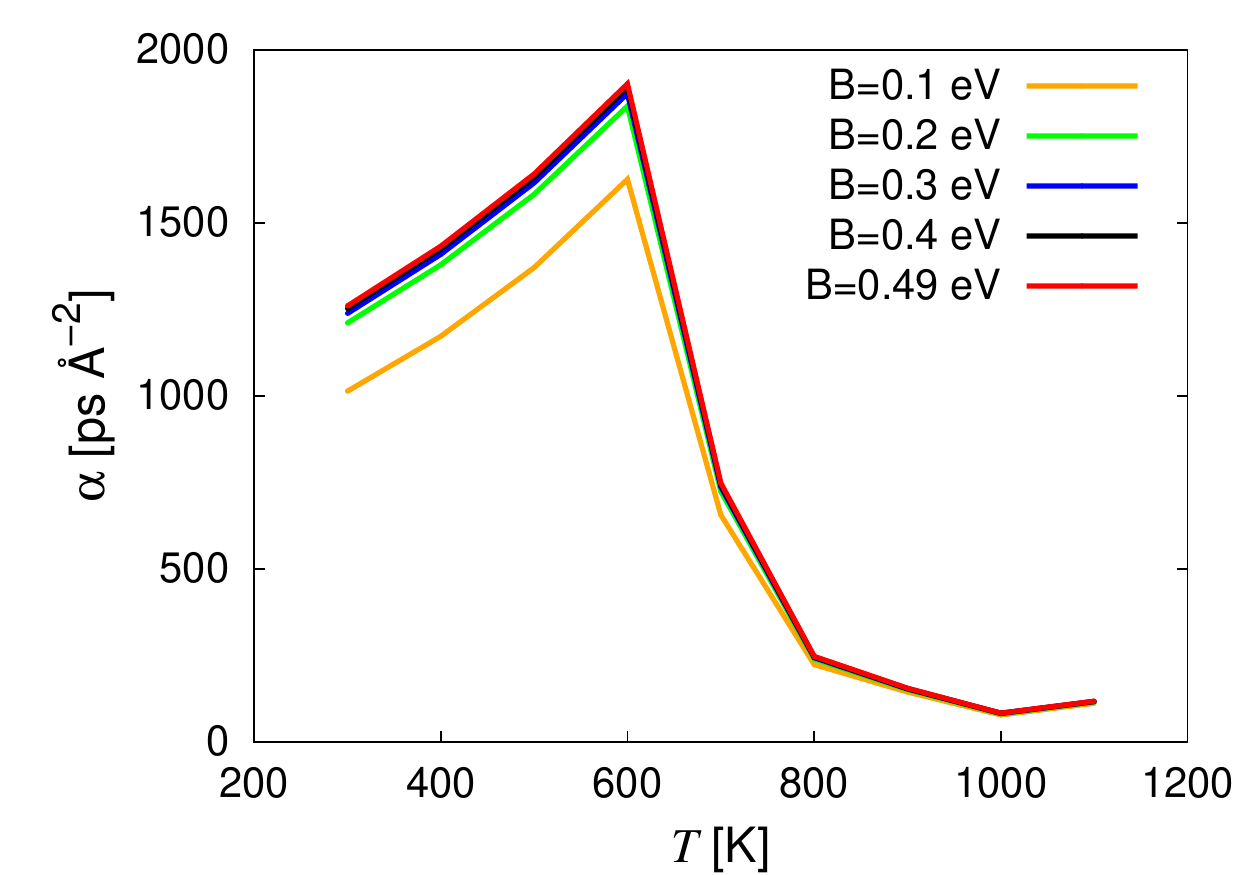}}
    \label{Fig:alpha_temperature_110}
}
\subfigure[]{
 {\includegraphics[width=0.35\textwidth]{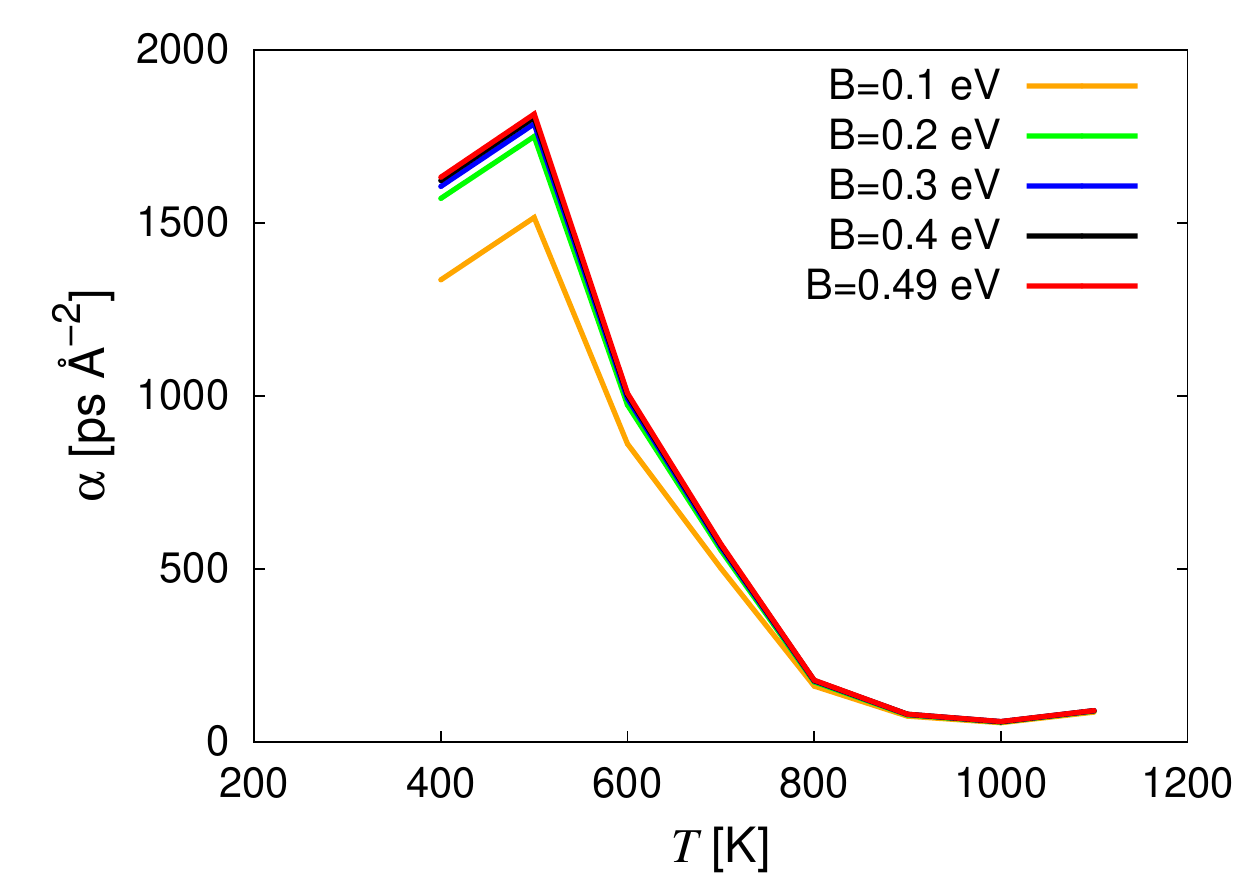}}
    \label{Fig:alpha_temperature_100}
}
\caption{Phase field parameter $\alpha$ as a function of temperature for several values of the free energy barrier $B$: \subref{Fig:alpha_temperature_111} (111) surface, \subref{Fig:alpha_temperature_110} (110) surface and \subref{Fig:alpha_temperature_100} (100) surface}
\label{Fig:alpha_temperature}
\end{figure}

Equations (\ref{Eq:m}) and (\ref{Eq:tau}) provide a direct recipe for computing $m(T,\theta)$ and  $M(T,\theta)$ from the functions $I_v(\Delta \bar{g})$ and $I_g(\Delta \bar{g})$. These are shown in Figs.~\ref{Fig:m_Temperature} and \ref{Fig:Mobility_Temperature} for different values of $B$. Equivalently, the values of $\epsilon (T,\theta)$ and $\alpha(T,\theta)$, defined in Eqs.~(\ref{Eq:epsilon}) and (\ref{Eq:alpha}), are shown in Figs.~\ref{Fig:epsilon_temperature} and \ref{Fig:alpha_temperature} respectively.
As the figures and equations show, the phase field parameters can be adjusted to deliver the appropriate interface energy and front velocity at constant temperature for an arbitrary value of the free energy barrier $B(T)$. In view of this observation, and the computational cost associated with the computation of $B(T)$, the value obtained in Sec.~\ref{Sec:Results_MD} is used. Other strategies for selecting $B$ based solely on computational cost, i.e.\ fixing the interface thickness, may also be used. In such case, the velocity and free energy of the interface would be exactly captured by the phase field model with a finite and potentially thick interface.  

\section{Simulations of Ge-nanocrystallization} \label{Sec:Discussions}
Next, the multiscale model developed in the previous sections is used to examine the interplay between the nucleation and growth mechanisms in the crystallization process.
For verification purposes, first we examine the ideal case of isotropic growth and demonstrate its consistency with the Avrami relations. To that end, we consider a temperature of 1100 K, nucleation rates that span several orders of magnitude, and growth conditions dictated by the phase field parameters previously computed for the three surface orientations. In these isotropic calculations, the parameter $\theta$ is simply a differentiating property for each growth condition. 

We begin by rendering Eq.~(\ref{Eq:phi_constantT}) dimensionless with the following characteristic length and timescale
\begin{equation}
\begin{split}
&l_c = R_c, \\
&t_c = \alpha R_c^2, \\
\end{split}
\end{equation}
where $R_c$ is the radius of the critical nuclei, given by Eq.~(\ref{Eq:CriticalRadius}). This specific choice of length and time scales has the property of delivering non-dimensional parameters $\epsilon$ and $\alpha$ that are independent of $\theta$. In particular, using Eq.~(\ref{Eq:m}), 
\begin{equation}
\begin{split}
&\bar{\epsilon}:= \frac{\epsilon}{R_c} = \frac{m(T,\theta) \Delta g(T)}{\sqrt{B} 2 g_{int}(T,\theta)}= \frac{\Delta \bar{g}(T)}{2 I_g(\Delta \bar{g}(T))}, \\
&\bar{\alpha} := \frac{\alpha R_c^2}{t_c} = 1. 
\end{split}
\end{equation}

\begin{figure}[ht]
\centering
\subfigure[]{
 {\includegraphics[width=0.35\textwidth]{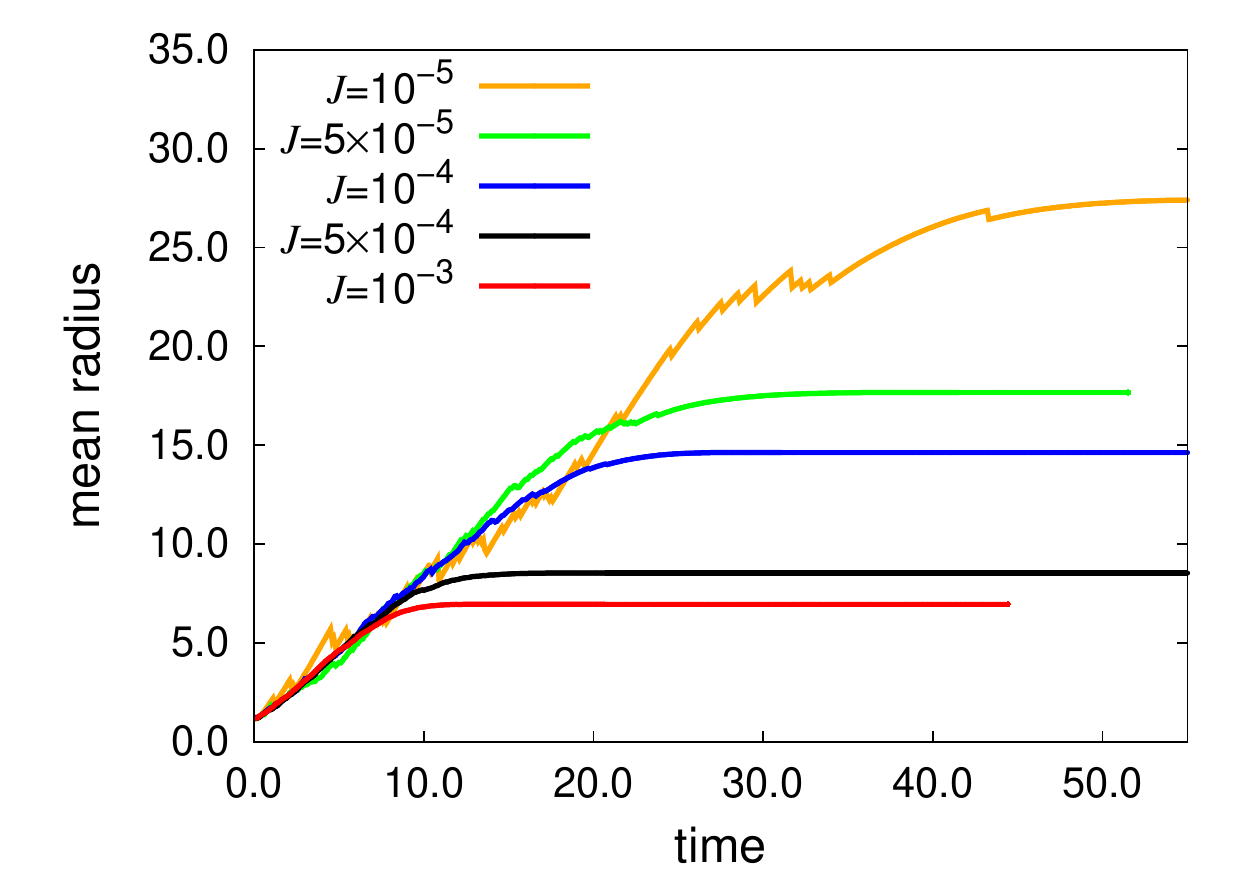}}
    \label{Fig:RadiusEvolution_adimensional}
}
\subfigure[]{
  {\includegraphics[width=0.35\textwidth]{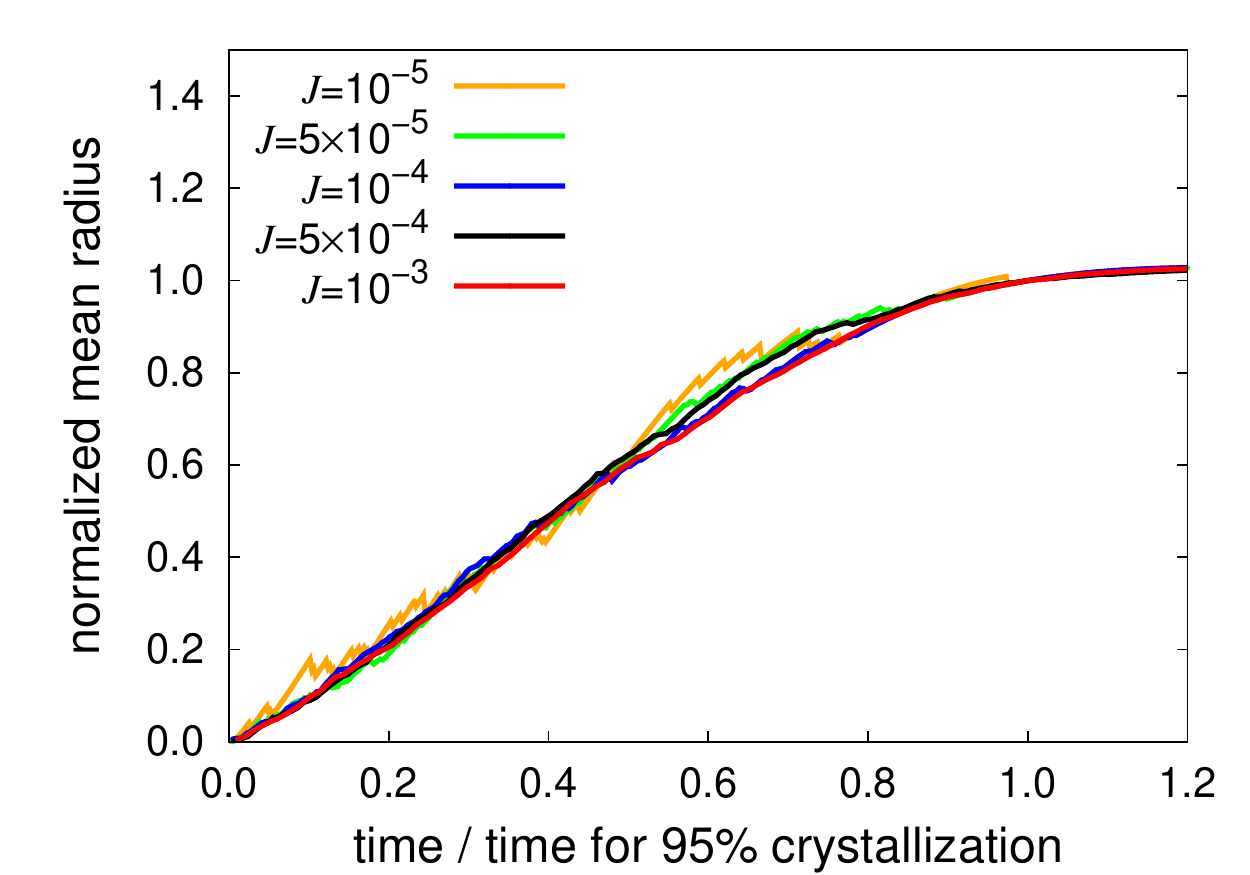}}
    \label{Fig:RadiusEvolution_adim_normalized}
}
\caption{Mean radius evolution with time for several nucleation rates. Isotropic growth conditions and spatially homogenous nucleation. \subref{Fig:RadiusEvolution_adimensional} Data non-dimensionalized using $l_c=R_c$ and $t_c =\alpha R_c^2$ as length and time scale respectively, \subref{Fig:RadiusEvolution_adim_normalized} Mean radius and time normalized with the corresponding value at $95\%$ crystallization. Simulation box: $400 \times 400$. Parameters of the simulation: $\bar{\epsilon} = 0.1807$, $\bar{\alpha } = 1$, $h=0.04$, $\Delta t = 0.001$.}
\end{figure}

\begin{figure}[ht]
\centering
\subfigure[]{
 {\includegraphics[width=0.35\textwidth]{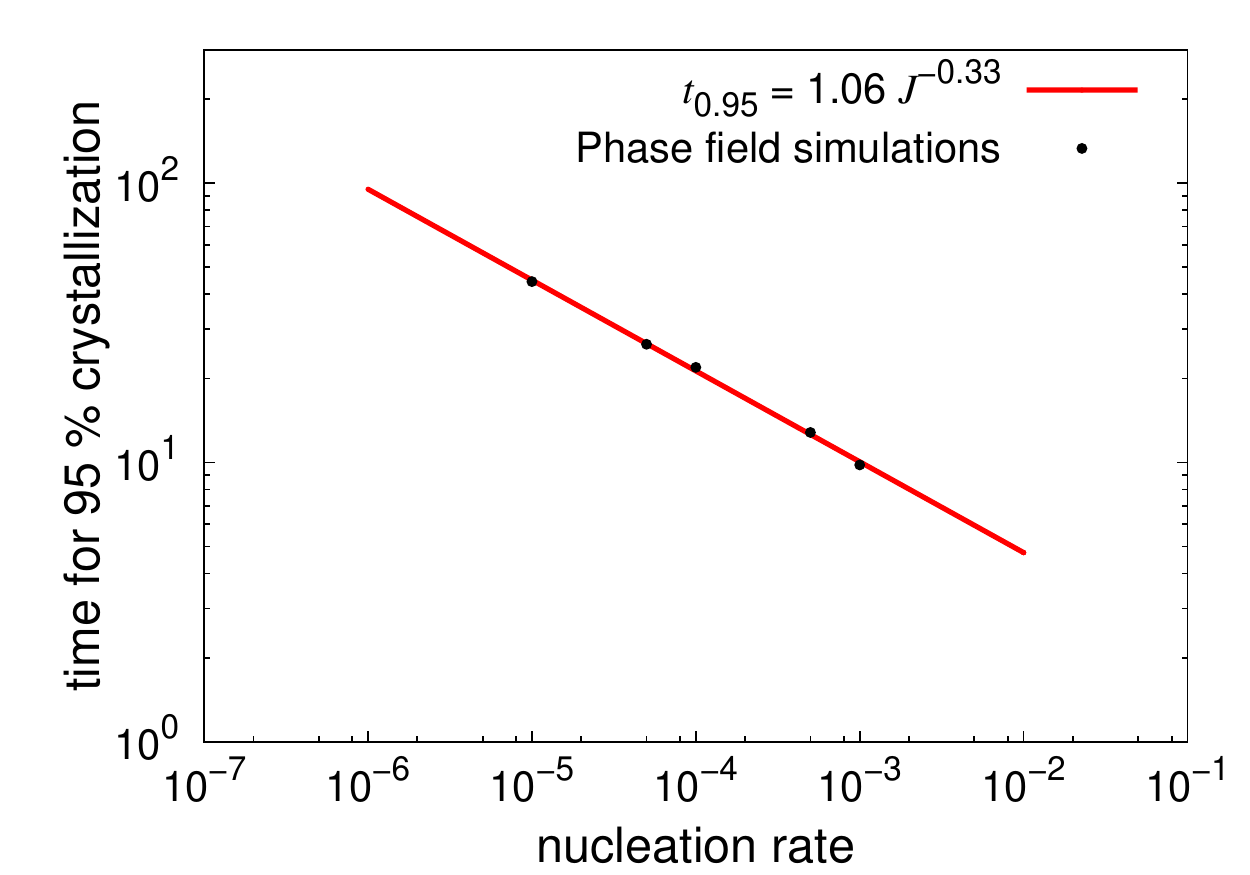}}
    \label{Fig:Time_095_J_adim}
}
\subfigure[]{
  {\includegraphics[width=0.35\textwidth]{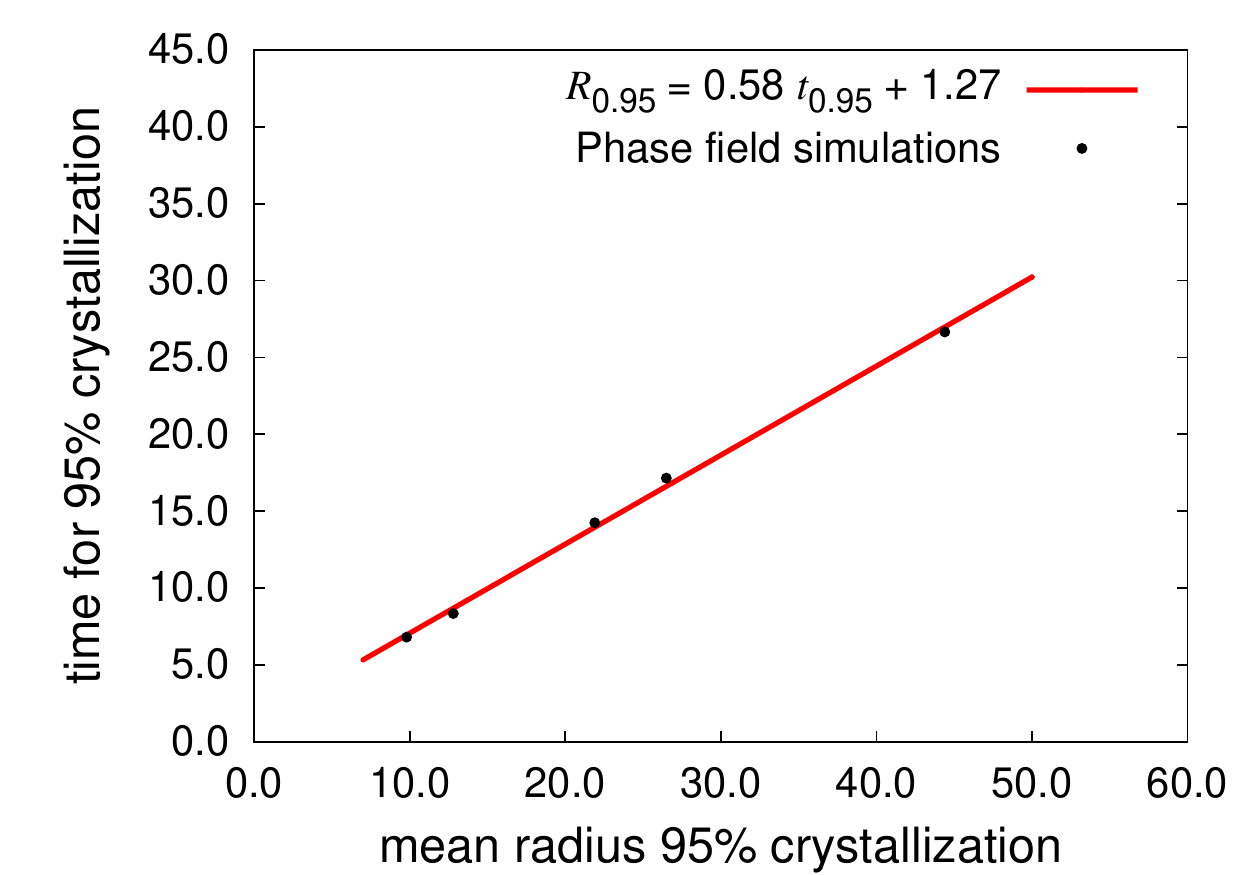}}
    \label{Fig:Radius_095_J_adim}
}
\caption{Scaling laws between the nucleation rate and the time and mean radius to achieve $95\%$ crystallization. Spatially homogeneous nucleation and isotropic growth conditions. Data non-dimensionalized with $l_c=R_c$ and $t_c =\alpha R_c^2$ as length and time scale respectively.}
\end{figure}

Consequently, the resulting non-dimensional phase field equation
\begin{equation}
\bar{\epsilon}^2 \frac{\partial\bar{\phi}}{\partial \bar{t}} = \bar{\epsilon}^2 \bar{\nabla} \bar{\phi} - g'_{dw}(\bar{\phi}) - \Delta \bar{g} \ q'(\bar{\phi})
\end{equation}
allows to examine all growth conditions with a single equation, which we proceed to couple with an explicit nucleation strategy, with non-dimensional nucleation rates ranging between $10^{-5}$ and $10^{-3}$. The numerical results attendant to this coupled nucleation and growth process are shown in Fig.~\ref{Fig:RadiusEvolution_adimensional}, where the time evolution of the mean radius during the nanocrystallization process is represented. These curves may be normalized with the time and mean radius at $95\%$ crystallization, delivering a universal crystallizatian curve, as shown in Fig.~\ref{Fig:RadiusEvolution_adim_normalized}. For their part, the time required to complete $95\%$ of the crystallization, and the mean radius at that stage of crystallization, satisfy very simple relations with the nucleation rate as depicted in  Figs.~\ref{Fig:Time_095_J_adim} and \ref{Fig:Radius_095_J_adim}. The slope of the fit to the numerical points in both curves coincide with the exponents of Avrami's relations, as detailed in Appendix \ref{Sec:Appendix_B}.

%

\begin{figure}
\begin{center}
    {\includegraphics[width=0.45\textwidth]{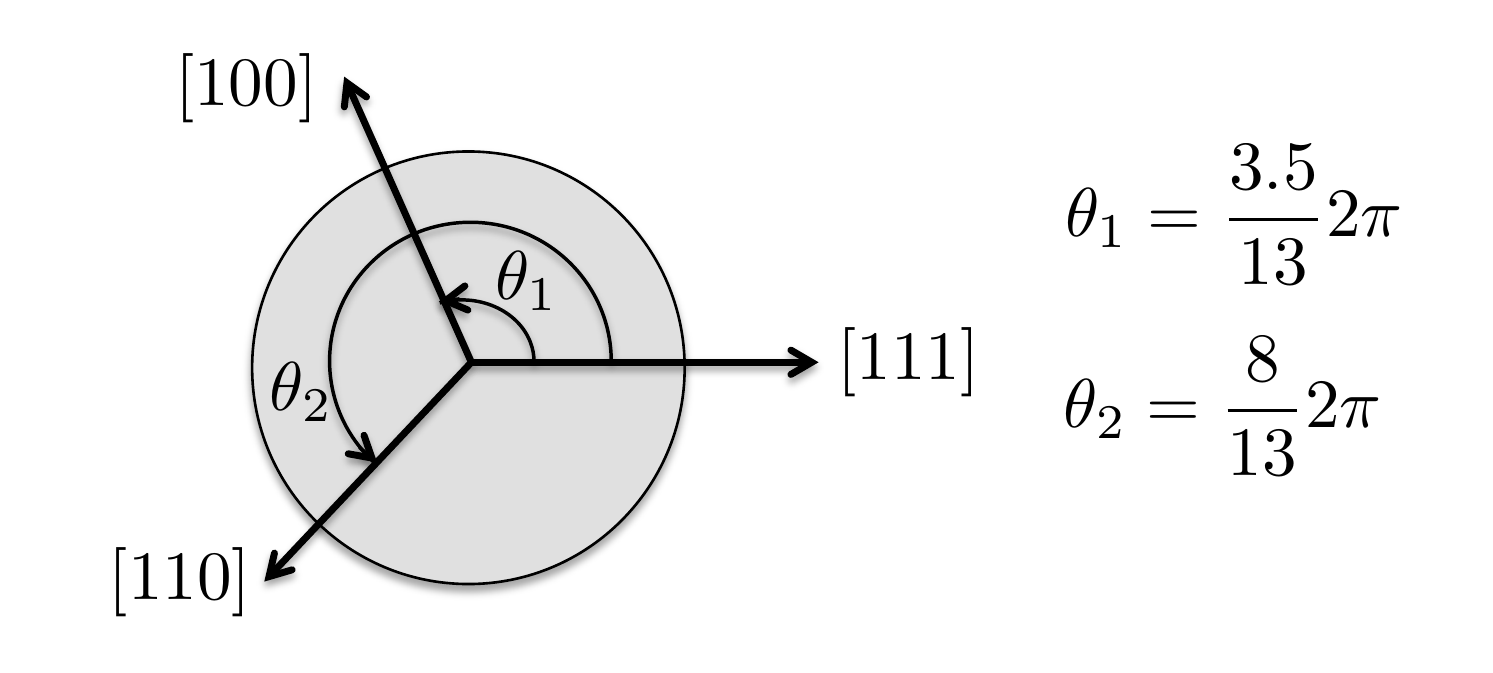}}
    \caption[]{The values of $\epsilon$ and $\alpha$ for a given misorientation $\theta \in [0, 2 \pi)$ will be obtained by linear interpolation of the values for orientations [111], [100] and [110] spaced as shown in the figure.}
    \label{Fig:Theta_interpolation}
\end{center}
\end{figure}

Next, we proceed to examine the case of anisotropic growth at $T=1100$ K using the surface orientation dependent data for the phase field parameters. In particular, we use a linear interpolation scheme between the $[111]$, $[100]$ and $[110]$ data previously computed, to define $\alpha(\theta)$ and $\epsilon (\theta)$ for an arbitrary value of the misorientation angle $\theta \in [0, 2 \pi)$, c.f.~Fig.~\ref{Fig:Theta_interpolation}. The specific spacing between the three data points has been chosen proportional to the number of distinct $\langle111\rangle$, $\langle100\rangle$ and $\langle110\rangle$ orientations in a diamond cell structure (4, 3 and 6 respectively).

\begin{figure}
\begin{center}
    {\includegraphics[width=0.45\textwidth]{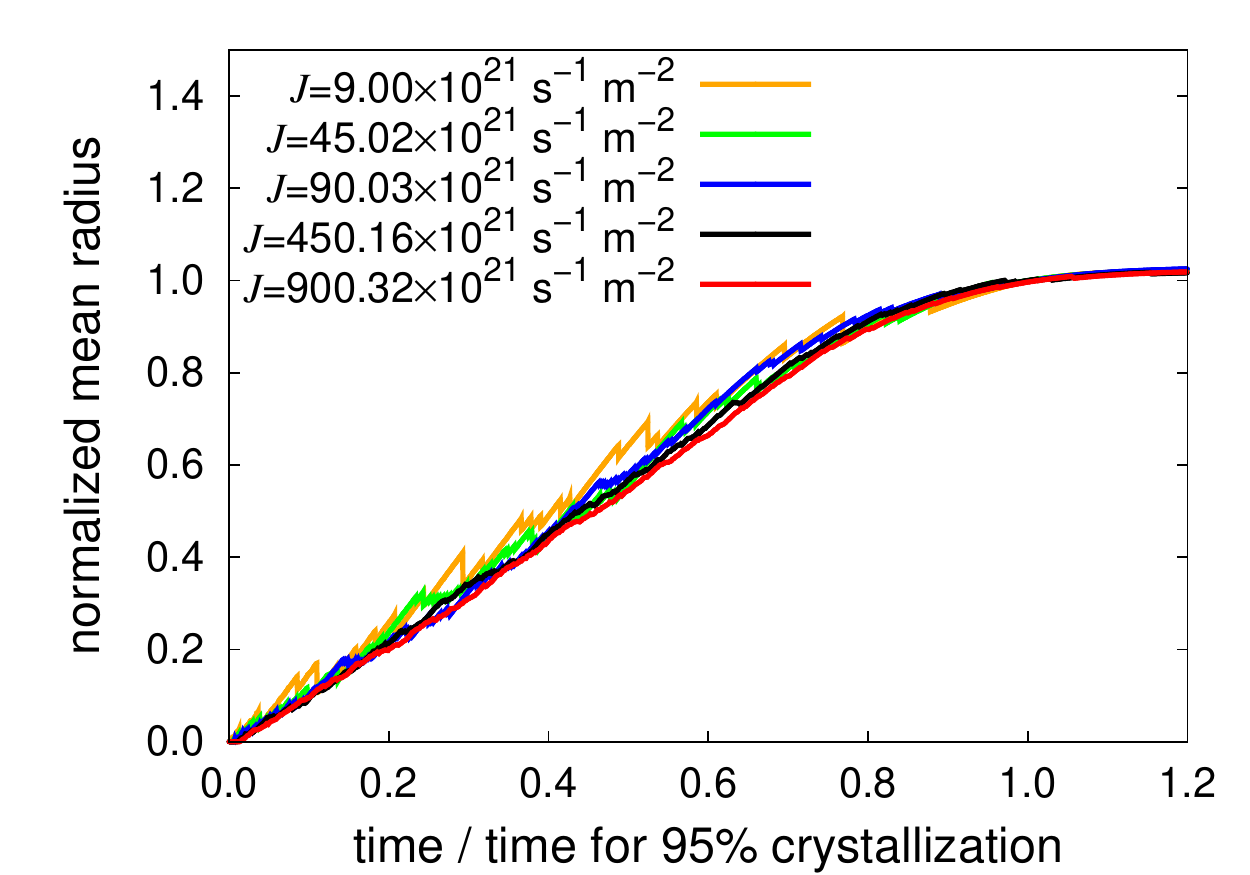}}
    \caption[]{Normalized mean radius evolution versus normalized time for several nucleation rates. Spatially homogeneous nucleation and anisotropic growth conditions.}
    \label{Fig:TR_anisotropic}
\end{center}
\end{figure}

\begin{figure}[ht]
\centering
\subfigure[]{
 {\includegraphics[width=0.35\textwidth]{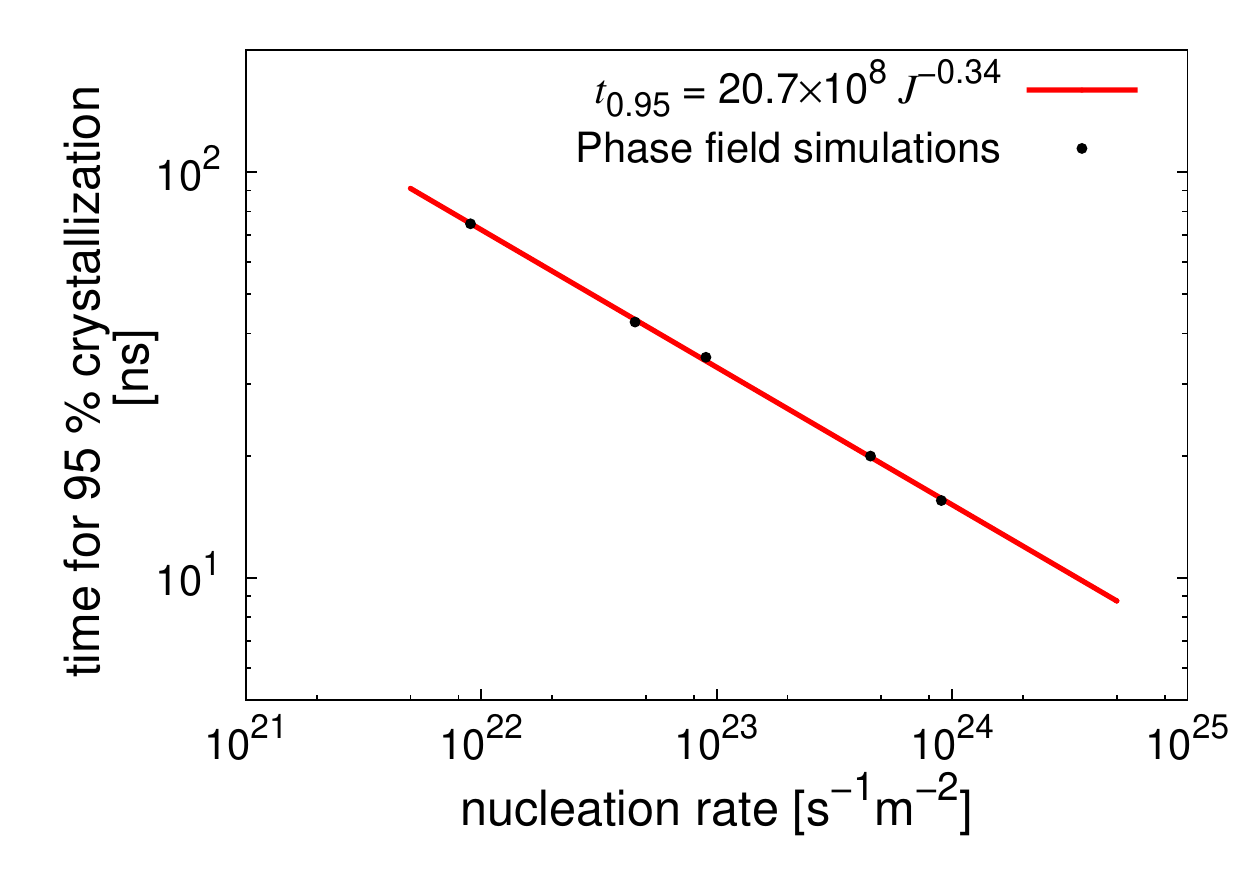}}
    \label{Fig:JT_anisotropic}
}
\subfigure[]{
  {\includegraphics[width=0.35\textwidth]{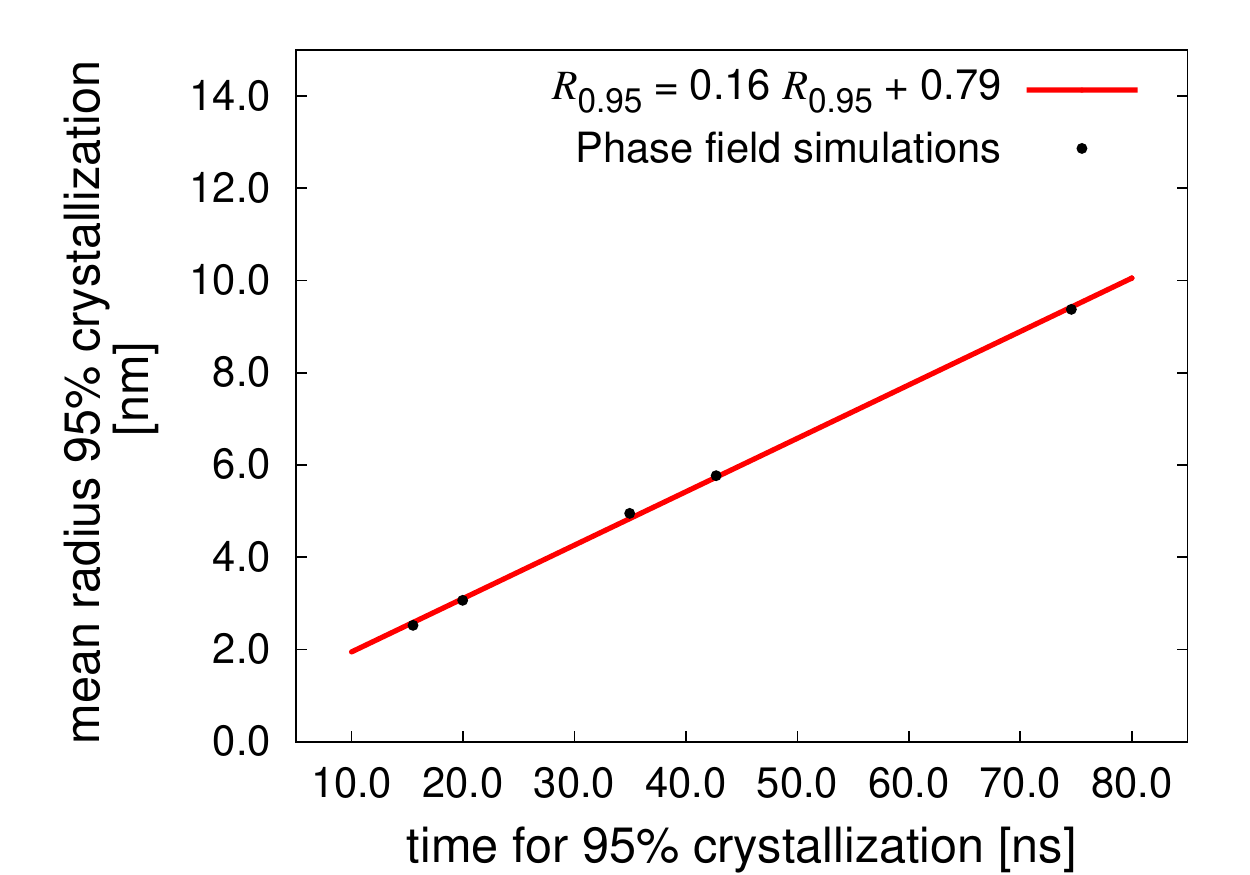}}
    \label{Fig:JR_anisotropic}
}
\caption{Scaling laws between the nucleation rate and the time and mean radius to achieve $95\%$ crystallization. Spatially homogeneous nucleation and anisotropic growth conditions.}
\end{figure}

%

\begin{figure}[ht]
\centering
\subfigure[]{
 {\includegraphics[width=0.35\textwidth]{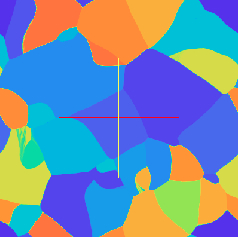}}
    \label{Fig:Final_Microstructure_J_L_400_1_omp}
}
\subfigure[]{
  {\includegraphics[width=0.35\textwidth]{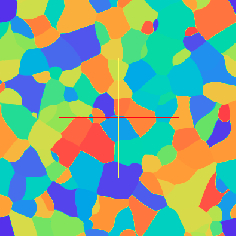}}
    \label{Fig:Final_Microstructure_J_L_400_3_omp}
}
\subfigure[]{
 {\includegraphics[width=0.35\textwidth]{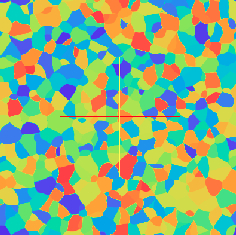}}
    \label{Fig:Final_Microstructure_J_L_400_5_omp}
}
\caption{Final nano-crystalline structure induced by spatially homogeneous nucleation and anisotropic growth conditions: \subref{Fig:Final_Microstructure_J_L_400_1_omp} $J=9\times10^{21}$, \subref{Fig:Final_Microstructure_J_L_400_3_omp} $J=90\times10^{21}$ and \subref{Fig:Final_Microstructure_J_L_400_5_omp} $J=900\times10^{21}$ s$^{-1}$~m$^{-2}$. Each color corresponds to a given crystallographic orientation, which is randomly assigned at nucleation.}
\label{Fig:Final_Microstructure_J_anisotropic}
\end{figure}

\begin{figure}
\begin{center}
    {\includegraphics[width=0.45\textwidth]{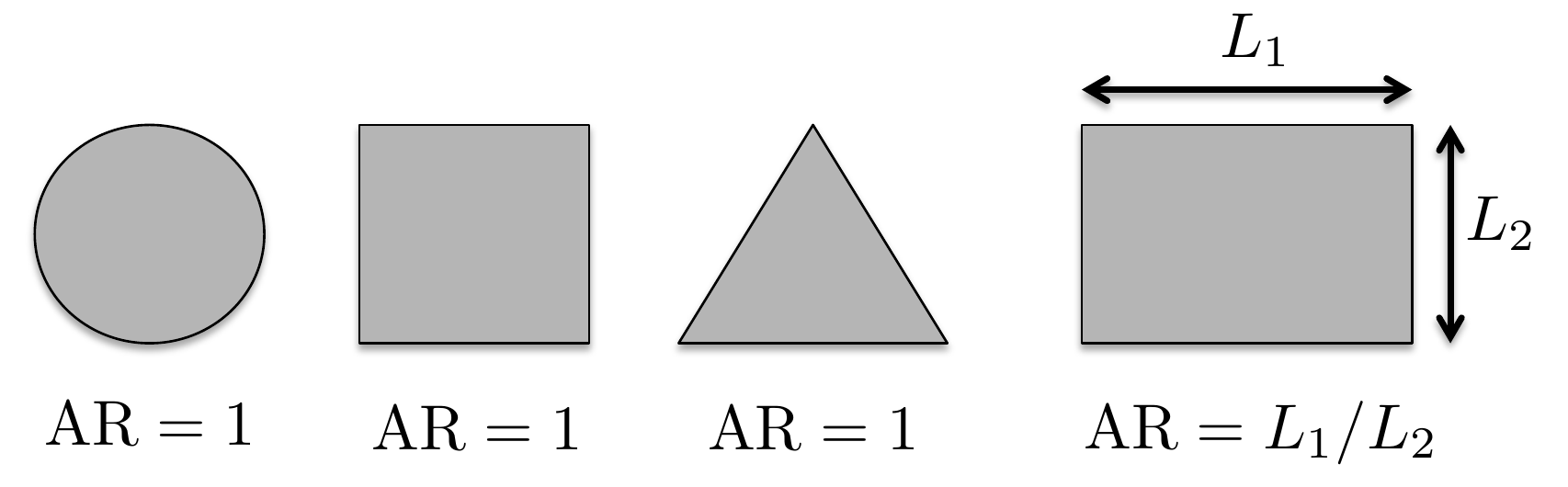}}
    \caption[]{Anisotropy ratio (AR) for several illustrative shapes.}
    \label{Fig:AR}
\end{center}
\end{figure}

\begin{figure}[ht]
\centering
\subfigure[]{
 {\includegraphics[width=0.45\textwidth]{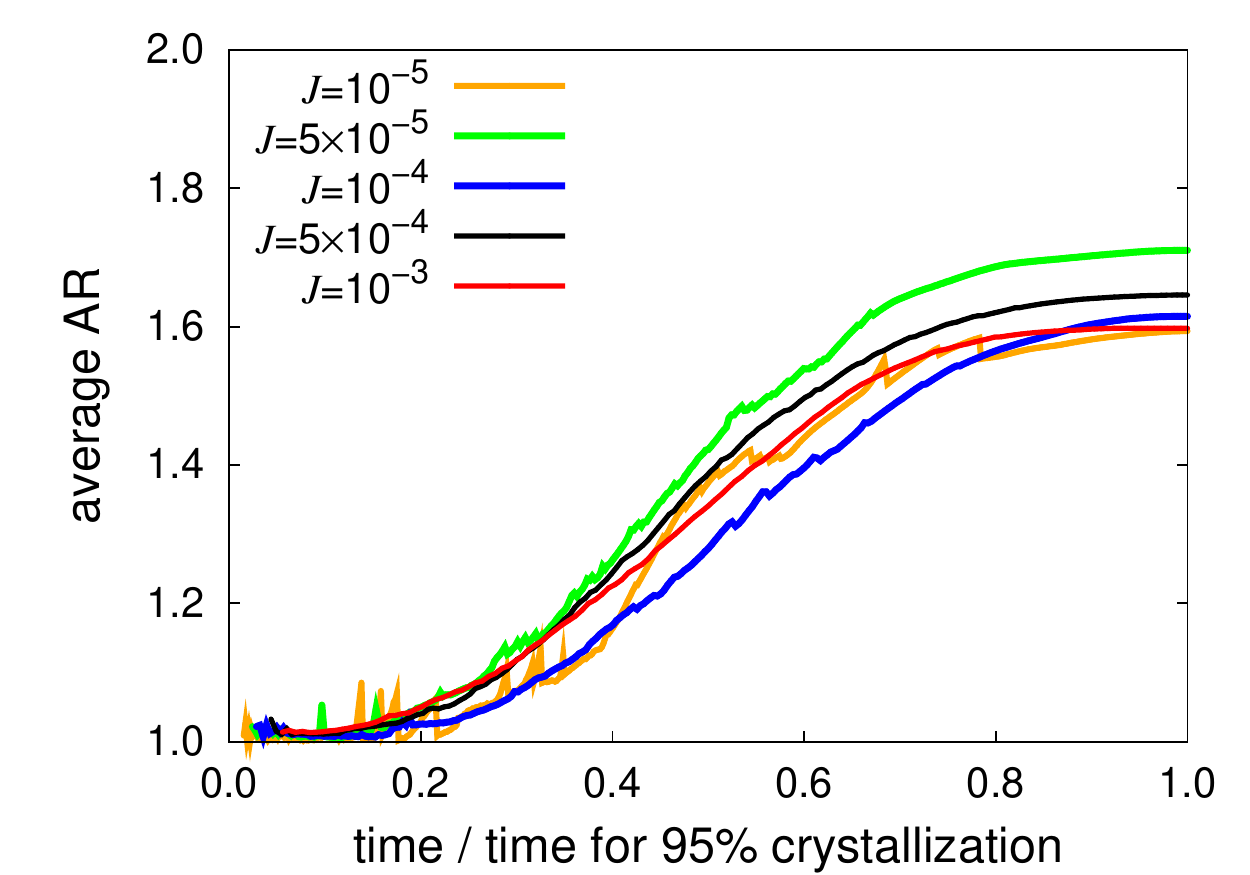}}
    \label{Fig:AR_isotropic_case_normalized}
}
\subfigure[]{
  {\includegraphics[width=0.45\textwidth]{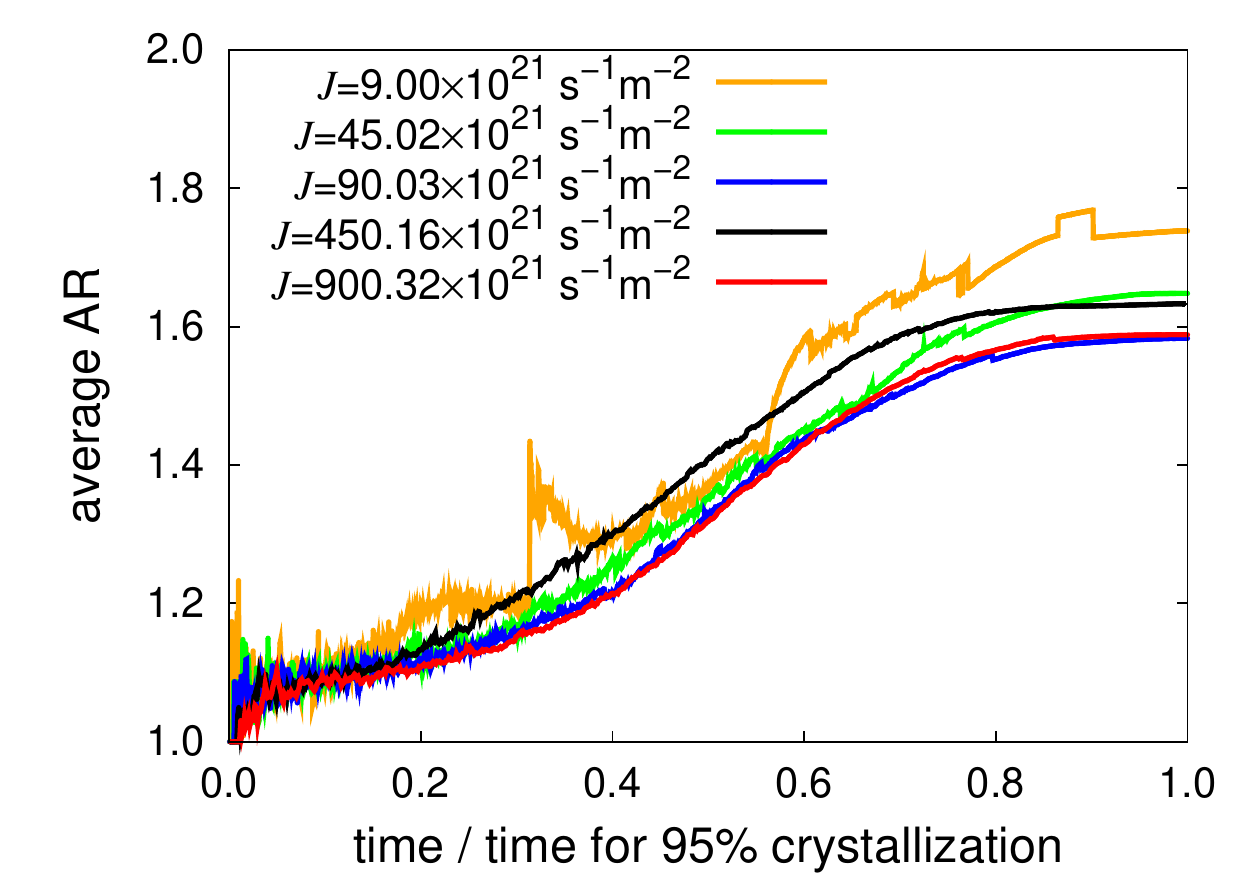}}
    \label{Fig:AR_anisotropic_case_normalized}
}
\caption{Time evolution of the average anisotropy ratio for the cases of: \subref{Fig:AR_isotropic_case_normalized} isotropic growth and \subref{Fig:AR_anisotropic_case_normalized} anisotropic growth.}
\label{Fig:AR_evolution}
\end{figure}

The resulting curves for the mean radius evolution as a function of time for different non-dimensional nucleation rates follow in a very similar manner to the isotropic case, c.f.~Fig.~\ref{Fig:TR_anisotropic}. Despite the strong growth anisotropy exhibited by Ge, c.f.~Sec.~\ref{Sec:Results_MD}, the time for crystallization and the mean radius of the final microstructure also follow fairly closely Avrami's relations, c.f.~Fig.~\ref{Fig:JT_anisotropic} and \ref{Fig:JR_anisotropic}. The physical reasoning behind this result lies on the fact that the final microstructure does not exhibit grains with a large aspect ratio, c.f.~Fig.~\ref{Fig:Final_Microstructure_J_anisotropic}.. The homogenous nucleation creates constraints to the growth of preexisting nuclei that are, on average, equi-spaced, leading to anisotropic ratios of the final microstructure similar to the simulations of isotropic growth, c.f. Fig.~\ref{Fig:AR_evolution}. The measure of anisotropy used in these calculations is
\begin{equation}
\text{AR} = \sqrt{\frac{\lambda_2}{\lambda_1}},
\end{equation}
where $\lambda_2 > \lambda_1$ are the eigenvalues of the gyration tensor of each grain, see Fig.~\ref{Fig:AR} for several example of AR.

\section{Summary and concluding remarks} \label{Sec:Summary}
The results shown in this paper indicate that the Ge nanocrystalline structure resulting from highly anisotropic growth and homogenous nucleation is fairly isotropic in the range of nucleation rates examined. As a result, the relations between the crystallization time, final average mean radius and nucleation rate follow fairly closely Avrami's relations. These relations could be potentially used to extract nucleation rates from experimental measurements of time-to-crystallization and grain size or vice versa. However, it must be kept in mind that the nucleation rates used here for numerical convenience may differ from those found in laboratory experiments depending on the experimental conditions \cite{Nikolova2010,germain1979}, and any exercise of extrapolation for the anisotropic curves is not backed by our model.

The numerical strategy chosen to obtain these results consists on phase field simulations that exactly deliver the interface kinetics and energetics obtained via atomistic calculations. This exact mapping between the atomistic and continuum solutions is based on two invariance relations of the Allen-Cahn equations at constant temperature. More precisely, we show in this paper that the non-dimensional steady state velocity and interfacial free energy of planar crystallization fronts satisfy universal relations with the non-dimensional free energy difference between the crystalline and amorphous phase.  These two relations allow for an explicit parametrization of the phase field equations with the atomistic data calculated from molecular dynamic simulations in suitably constructed subsystems using state-of-the-art interatomic potentials.

\appendix
\section{Free energy of planar still interface} \label{Sec:Appendix_A}
Consider a flat crystalline-amorphous interface with an energy landscape described by a perfect double-well potential, i.e. with $\Delta g = 0$. The equilibrium solution consists of a static interface that satisfies
\begin{equation}
 0 = m^2 \nabla^2 \phi - B \ g'_{dw}(\phi).
\end{equation}

For this specific case, the interface profile is also the minimizer of the interface free energy functional, c.f.~Eq.~(\ref{Eq:gint}),
\begin{equation} \label{Eq:g_int_still}
g_{int}=m^2 \Bigg[ \int_{\ell} \frac{(\nabla \phi)^2}{2} dx\Bigg]  + B \Bigg[ \int_{\ell}g_{dw}(\phi)dx\Bigg], 
\end{equation}
where $x$ to be the coordinate orthogonal to the interface. Then, the integrand of this functional can be seen as a Lagrangian, that exclusively depends on $\phi$ and $\frac{d\phi}{dx}$ and does not depend explicitly on the variable $x$,

\begin{equation}
L\left(\phi,\frac{d\phi}{dx}\right) = \frac{m^2}{2} (\nabla \phi)^2 + B\ g_{dw}(\phi).
\end{equation}

 As a result, by Noether's theorems, the corresponding Hamiltonian (Legendre transform of the Lagrangian) is independent of the variable $x$. More precisely,

\begin{equation}
L - \frac{\partial L}{\partial (d\phi/dx)}\ \frac{d\phi}{dx} =  \text{constant},
\end{equation}
or equivalently,
\begin{equation}
-\frac{m^2}{2} (\nabla \phi)^2 + B\ g_{dw}(\phi)=  \text{constant}.
\end{equation}

The constant may be evaluated far away from the interface, where $\phi$ is either $0$ or $1$, delivering a zero value. As a result, the interface energy is partitioned equally between the two terms in Eq.~(\ref{Eq:g_int_still}), 
\begin{equation} \label{Eq:Equipartition}
\frac{g_{int}}{2}=m^2 \Bigg[ \int_{\ell} \frac{(\nabla \phi)^2}{2} dx\Bigg]  = B \Bigg[ \int_{\ell}g_{dw}(\phi)dx\Bigg]. 
\end{equation}

The value of either of these two integrals may be evaluated for the equilibrium solution, c.f. Eq.~(\ref{Eq:Analytic_phi}), giving
\begin{equation}
g_{int}=\frac{\sqrt{8}}{3}\epsilon B\simeq\epsilon B.
\end{equation}

We further remark that the value of the integrals in Eq.~(\ref{Eq:Equipartition}) are not very sensitive to the specific interface profile. A simple piecewise linear profile with interface thickness $\epsilon$, i.e. $\phi(x)=\frac{x}{\epsilon}$ for $0<x<\epsilon$, would deliver
\begin{equation}
g_{int}=\frac{16}{15}\epsilon B\simeq\epsilon B.
\end{equation}

\section{Crystallization time and mean radius at crystallization as a function of the nucleation rate} \label{Sec:Appendix_B}
Avrami's equation \cite{Avrami1940} describes the kinetics of crystallization under the assumption of homogeneous nucleation, constant nucleation rate and constant growth rate for the radius of each nuclei independently of its size. For a two dimensional system, it reads
\begin{equation}
\log \left( - \log (1-f_C)\right) = \log \left(\frac{\pi \dot{R}^2}{3} J \right) + 3 \log t,
\end{equation}
where $f_C$ is the volume fraction of the crystalline phase and $\dot{R}$ is the growth rate. This equation immediatly delivers the relation between the time required to acheive $95\%$ crystallization $(f_C=0.95)$ and the nucleation rate. Namely, it is of the form
\begin{equation}
t_{0.95} = C J^{-1/3},
\end{equation}
where $C=\left(\frac{\pi \dot{R}^2}{3} \right)^{-1/3} \left(-\log 0.05\right)^{1/3}$.

On its side, the mean radius at $95\%$ crystallization ($\bar{R}_{0.95}$) satisfies
\begin{equation}
\frac{0.95 V}{N_{0.95}} = \pi \bar{R}_{0.95}^2,
\end{equation} 
where $V$ is the total volume and $N_{0.95}$ is the number of nuclei composing the nanocrystalline structure. The later, can be computed as
\begin{equation} \label{Eq:N}
\begin{split}
N_{0.95} &= \int_0^{t_{0.95}} \dot{N} \, dt =  \int_0^{t_{0.95}} J V \left(1-f_C(t) \right) \, dt \\
&= JV \int_0^{CJ^{-1/3}}\exp \left(-\frac{\pi \dot{R}^2}{3} t^3 J \right) \, dt.
\end{split}
\end{equation}

By use of the change of variables $\tau = t J^{1/3}$, Eq.~\ref{Eq:N} simplifies to
\begin{equation}
N_{0.95} = J^{2/3} V \int_0^C \exp \left( -\frac{\pi \dot{R}^2}{3} \tau^3\right) \, d\tau = C_2 J^{2/3},
\end{equation}
with $C_2$ solely dependent on $\dot{R}$. Then, the mean radius at crystallization can be expressed as
\begin{equation}
\bar{R}_{0.95} = \left(\frac{0.95 V}{\pi C_2}\right)^{1/2} J^{-1/3}
\end{equation}
or
\begin{equation}
\bar{R}_{0.95} =  \left(\frac{0.95 V}{\pi C^2 C_2}\right)^{1/2} t_{0.95}.
\end{equation}

\section*{Acknowledgements}
This work performed under the auspices of the U.S. Department of Energy by Lawrence Livermore National Laboratory under Contract DE AC52-07NA27344. C. Reina acknowledges support from the Lawrence Fellowship program at Lawrence Livermore National Laboratory. J. Marian acknowledges support from DOE's Early Career Research Program.
\bibliographystyle{unsrt}

\end{document}